%% file: mr_opt_response_change_model_12_arxiv.tex
\documentclass[%
reprint,
superscriptaddress,
amsmath,amssymb,
prl,
noeprint,
longbibliography,
]{revtex4-2}

\usepackage{graphicx}
\graphicspath{{./fig/}}

\usepackage{dcolumn}
\usepackage{bm}
\usepackage{physics}
\usepackage{siunitx}
\usepackage{comment}
\usepackage{hyperref}
\usepackage{threeparttable}

\newcommand{\beginsupplement}{%
	\setcounter{table}{0}%
	\renewcommand{\thetable}{S\arabic{table}}%
	
	\setcounter{figure}{0}%
	\renewcommand{\thefigure}{S\arabic{figure}}%
	
	\setcounter{equation}{0}%
	\renewcommand{\theequation}{S\arabic{equation}}%
}
\makeatletter
\newcommand{\SuppressMainTOC}{%
	\let\orig@addcontentsline\addcontentsline
	\renewcommand{\addcontentsline}[3]{}
}
\newcommand{\RestoreTOC}{%
	\let\addcontentsline\orig@addcontentsline
}
\makeatletter

\begin{document}

\SuppressMainTOC
	
\preprint{APS/123-QED}

\title{Spatially Resolved Optical Responses of \\a Superconducting Nanowire Microwave Resonator}

\author{Rento Hirotsuru}
\thanks{These two authors contributed equally}

\address{Department of Physics, Graduate School of Engineering Science, Yokohama National University, 79-5 Tokiwadai, Hodogaya, Yokohama, Kanagawa, 240-8501, Japan}

\author{Hodaka Kurokawa}
\thanks{These two authors contributed equally}
\email[E-mail: ]{kurokawa-hodaka-hm@ynu.ac.jp}
\address{Quantum Information Research Center, Institute of Advanced Sciences, Yokohama National University, 79-5 Tokiwadai, Hodogaya, Yokohama, Kanagawa, 240-8501, Japan}
\author{Kazuyo Takaki}
\address{National Institute of Information and Communications Technology, 588-2 Iwaoka, Nishi-ku, Kobe, Hyogo, 651-2492, Japan}
\author{Hirotaka Terai}
\address{National Institute of Information and Communications Technology, 588-2 Iwaoka, Nishi-ku, Kobe, Hyogo, 651-2492, Japan}
\author{Hideo Kosaka}
\email[E-mail: ]{kosaka-hideo-yp@ynu.ac.jp}
\address{Department of Physics, Graduate School of Engineering Science, Yokohama National University, 79-5 Tokiwadai, Hodogaya, Yokohama, Kanagawa, 240-8501, Japan}
\address{Quantum Information Research Center, Institute of Advanced Sciences, Yokohama National University, 79-5 Tokiwadai, Hodogaya, Yokohama, Kanagawa, 240-8501, Japan}

\begin{abstract}
Understanding the optical response of a superconducting microwave resonator is crucial for applications ranging from single-photon detection to quantum transduction between the microwave and optical domains. This topic  is gaining significant attention for scaling up quantum computers. However, interactions between the  pump light and superconducting resonators often induce unintended resonance frequency shifts and linewidth broadening.
In this study, we measure  the local optical response of a NbTiN nanowire resonator using a laser-scanning microwave spectroscopy system integrated with a dilution refrigerator. The optical response of the resonator correlates with the resonance modes and position, which we attribute to interactions between two-level systems (TLSs) and laser-induced phonons. 
These findings establish the TLS-phonon interactions as the dominant mechanism underlying laser-induced resonance variations, providing critical insights for the design and optimization of quantum transducers and superconducting single-photon detectors. 
\end{abstract}

\maketitle


Superconducting microwave resonators are essential building blocks for modern quantum technologies. Their functions include the measurement of superconducting qubits \cite{Wallraff2004, Blais2021}, storage of bosonic microwave qubits \cite{Ni2023,Sivak2023,Ganjam2024,Putterman2024}, realization of strong coupling to artificial atoms in solid-state systems \cite{Mi2018,Samkharadze2018,Wang2023}, single-photon detection \cite{Vayonakis2003,Natarajan2012}, and quantum interfacing between microwave and optical photons \cite{Andrews2014,Mirhosseini2020,Jiang2022,Sahu2023a,vanThiel2025}. 
Among these applications, the interaction between light and superconducting circuits has gained significant attention for scaling up quantum computers via microwave--optical links. However, optical photons can also induce degradation in superconducting devices. Microwave--optical quantum transducers \cite{Han2020,Mirhosseini2020,Meesala2023,Jiang2022,Bartholomew2020,Weaver2022,Warner2023,Xie2024,Kurokawa2022,Dirnegger2025,Weaver2025} suffer from pump-light leakage during quantum frequency conversion \cite{Xu2022b,Weaver2022,Zhao2024}. Precise understanding of the optical response is crucial for the design and optimization of microwave resonators coexisting with optical ports including opto-mechanical systems \cite{Mirhosseini2020,Jiang2022,Weaver2022,Zhao2024}, electro-optical systems \cite{Xu2022b,Warner2023}, and optically addressable solid-state defects or atoms \cite{Tsuchimoto2022,Kurokawa2022,Xie2024} for quantum transduction. Furthermore, high-energy particle strikes (cosmic rays, $\gamma$ rays) induce long-range correlated errors in superconducting processors because of athermal phonons being propagated  \cite{Wilen2021,McEwen2022,Iaia2022,Harrington2024,Li2024a,McEwen2024,Anthony-Petersen2024}. Investigating  optical responses can reveal the interplay among high-energy particles, superconducting circuits, two-level systems (TLSs), and phonons.

This study investigates  the effects of local optical illumination on a superconducting microwave resonator, with a particular focus on illumination-position dependence. This approach applies a custom-built laser-scanning microwave spectroscopy system integrated with a dilution refrigerator. We employ  a high-impedance resonator, which is of growing interest in circuit quantum electrodynamics experiments \cite{Samkharadze2016,Niepce2019,Weaver2022,Jiang2022,Meesala2023} including quantum transduction, owing to its high zero-point fluctuation voltages. Furthermore, its high  sensitivity to optical power enables experiments at lower optical powers compared to a typical 50-$\Omega$ resonator.

We reveal that degradation of the quality factor, $Q$, and shifts in resonance frequency depend on two factors: the resonator mode and the illumination position. In particular, we find that the local electric field strength interacting with TLSs contributes significantly to the optical response. We estimate that optically induced nonequilibrium phonons cause changes in TLS populations, resulting in unintended shifts in the resonance spectrum. These findings provide critical insight for the design and operation of quantum transducers and superconducting single-photon detectors, the interaction between TLS and phonons, and 
the impacts of high-energy particle irradiation on superconducting circuits \cite{Benevides2024}, which are catastrophic to quantum processors \cite{Wilen2021,McEwen2022,Iaia2022,Harrington2024,Li2024a,McEwen2024,Anthony-Petersen2024}.

\par 
\begin{figure*}
	\centering
	\includegraphics[width=17cm,clip]{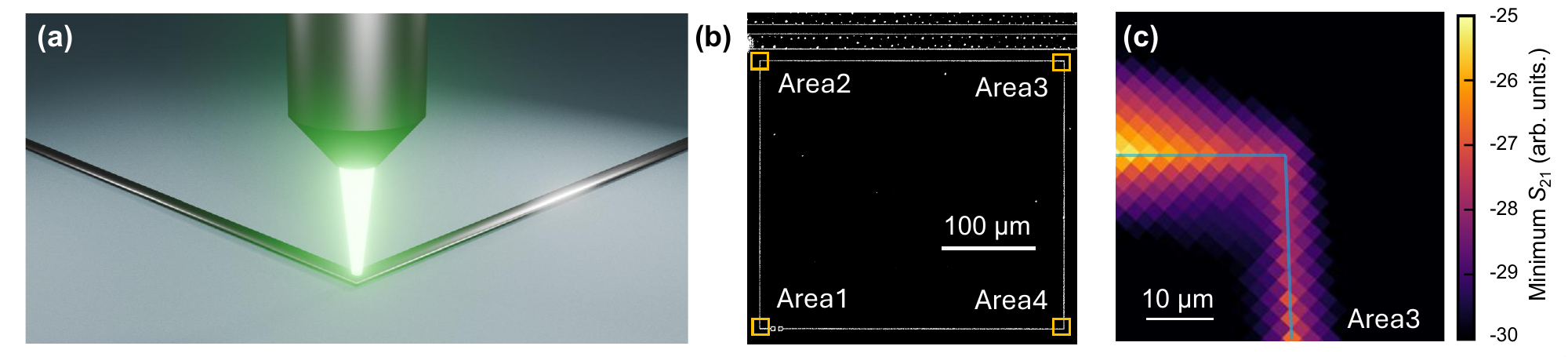}
	\caption{Spatially resolved optical illumination experiments. (a) Schematic of the optical illumination experiments. (b) Photograph of the resonator. Areas 1--4 correspond to the four corner of the resonator. (c)  Two-dimensional (2D) optical response map of the superconducting nanowire. The 2D optical response map of the resonator is constructed by sweeping the laser illumination position using the galvanometer mirror and extracting the minimum $S_{21}$ amplitude at each coordinate. The blue line probably  corresponds to the nanowire.
	}
	\label{Fig1}
\end{figure*}

Optical illumination experiments (Fig. \ref{Fig1}(a)) are performed using a custom-built laser-scanning optical system integrated with a dilution refrigerator at a base temperature of $\sim$15 mK. A 515-nm laser beam, focused on the sample surface using an objective lens, is scanned via a two-axis galvanometer mirror. The modulation of the transmitted microwave signal by the laser irradiation is measured using a vector network analyzer. Thus, the structure of the superconducting half-wavelength resonator--fabricated from a 10-nm-thick, 150-nm-wide NbTiN nanowire--is imaged (Fig. \ref{Fig1}(b)). The focused laser locally alters the resonance spectrum, allowing visualization of the nanowire, as shown in Fig. \ref{Fig1}(c).

To investigate the dependence on the resonance mode and illumination position, we measure the optical responses of $1/Q$ and $\Delta f_\mathrm{r}/f_\mathrm{r}$ across 16 combinations of four resonance modes (first, second, third, and fifth) and four illumination positions (corners of Areas 1--4) (Fig. \ref{fig3} (a)(b)). Here, we exclude the fourth mode due to its shallow resonance dip (see Supplementary Notes for details of the resonance spectrum). Additionally, the experiments are performed in the high-microwave power regime in which TLS losses saturate. 
As shown in Fig. \ref{fig3} (b), the changes in $1/Q$ and $\Delta f_\mathrm{r}/f_\mathrm{r}$ with optical power vary with the resonance mode and illumination position. $1/Q$ increases linearly with optical power for all modes and positions with different slopes, $\gamma$, where $\gamma$ is defined by $1/Q=\gamma P_\mathrm{opt}+1/Q_{0}$, where $P_\mathrm{opt}$ is the laser power. 
The optical responses in $\Delta f_\mathrm{r}/f_\mathrm{r}$ differ both in magnitude and sign. Notably, three patterns are observed: low-frequency shift (red shift), high-frequency shift (blue shift), and no significant change. Phenomenologically, $\Delta f_\mathrm{r}/f_\mathrm{r}$ can be fitted by $\Delta f_\mathrm{r}/f_\mathrm{r} = \delta_1 P_\mathrm{opt}-\delta_2[1-\mathrm{exp}(-\delta_3P_\mathrm{opt})]$, where the first term represents the linear frequency shift and the second term represents the saturating lower frequency shift. We focus on the linear term below and discuss possible reasons for nonlinear behavior in Supplementary Discussion.

A dominant factor contributing to changes in $1/Q$ and $\Delta f_\mathrm{r}/f_\mathrm{r}$ is  quasiparticles in the superconductor generated by the laser photons. The observed increase in $1/Q$ is possibly due to a rise in the quasiparticle density, $n_\mathrm{qp}$ ($\Delta(1/Q)\propto n_\mathrm{qp} \propto P_\mathrm{opt}$ \cite{Benevides2024}). 
The frequency shift caused by quasiparticle generation can be approximated as $\Delta f_\mathrm{r}/f_\mathrm{r} \simeq - \Delta n_\mathrm{qp}/(2n_\mathrm{s})$ (see Supplementary Notes). 
Thus, $f_\mathrm{r}$ shifts linearly toward lower frequencies with quasiparticle generation. However, the observed upward shifts in $\Delta f_\mathrm{r}/f_\mathrm{r}$ cannot be explained solely by the linear downward shift expected from quasiparticle generation, suggesting that additional mechanisms contribute.


Remarkably, the comparison between the blue and red shifts indicates that the degree of blue shift correlates with a larger slope in $1/Q$ (Fig. \ref{fig3} (b)). Additionally, the direction of frequency shifts correlates with the spatial distribution of current density in the nanowire for each resonance mode, as shown by the two-dimensional color-map inset in Fig. \ref{fig3} (b). When the illumination occurs in regions of low current density, a blue shift is observed, whereas a red shift occurs in regions with high current density. Although this trend does not hold for certain higher frequency modes, such as the third and fifth modes at Area 4, most cases are consistent with this assumption.

\begin{figure*}[t]
	\centering
	\includegraphics[width=17cm,clip]{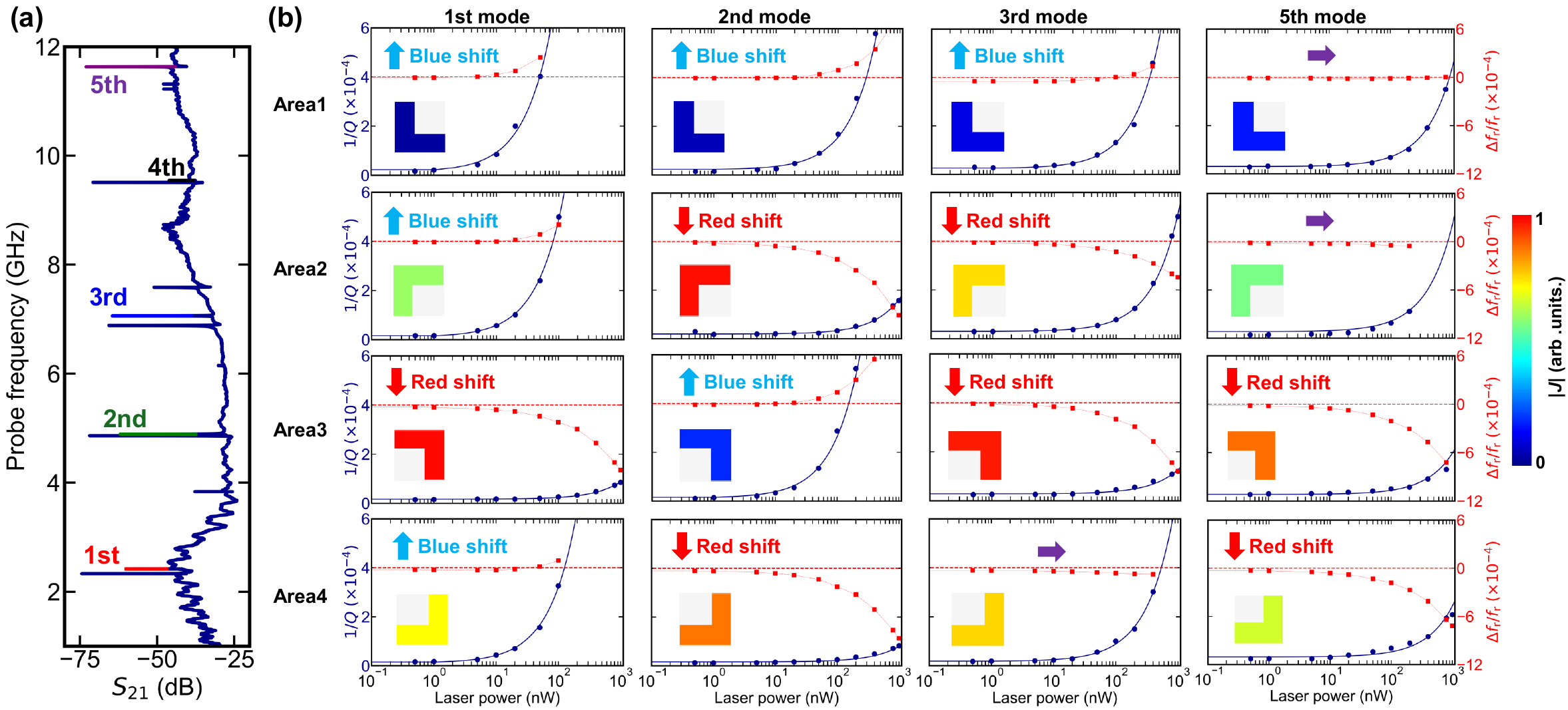}
	\caption{Optical response of the resonator across four modes and four positions. (a) $S_{21}$ spectrum of the resonator. The five resonance modes are labeled as the first through fifth modes. Additional resonance modes originate from other resonators located on the same chip. (b) Changes in $1/Q$ and $\Delta f_\mathrm{r}/f_\mathrm{r}$ under laser illumination are plotted for different modes and positions. $Q$ denotes the internal quality factor. 
	Fitting curves for $1/Q$ and $\Delta f_\mathrm{r}/f_\mathrm{r}$ are also shown. The horizontal red dashed line indicates the resonance frequency measured without optical illumination.
		Three  response types based on frequency shifts are: low-frequency shift (red shift), high-frequency shift (blue shift), and subtle shift (purple). Simulated current-density distributions for each position and mode are shown in the inset. The color is normalized by the maximum current density for each mode. The corners are enlarged to highlight the current density in each mode. To prevent current crowding at the corners, the current density is averaged, excluding the corners. }
	\label{fig3}
\end{figure*}

\begin{figure}[]
	\centering
	\includegraphics[width=8cm]{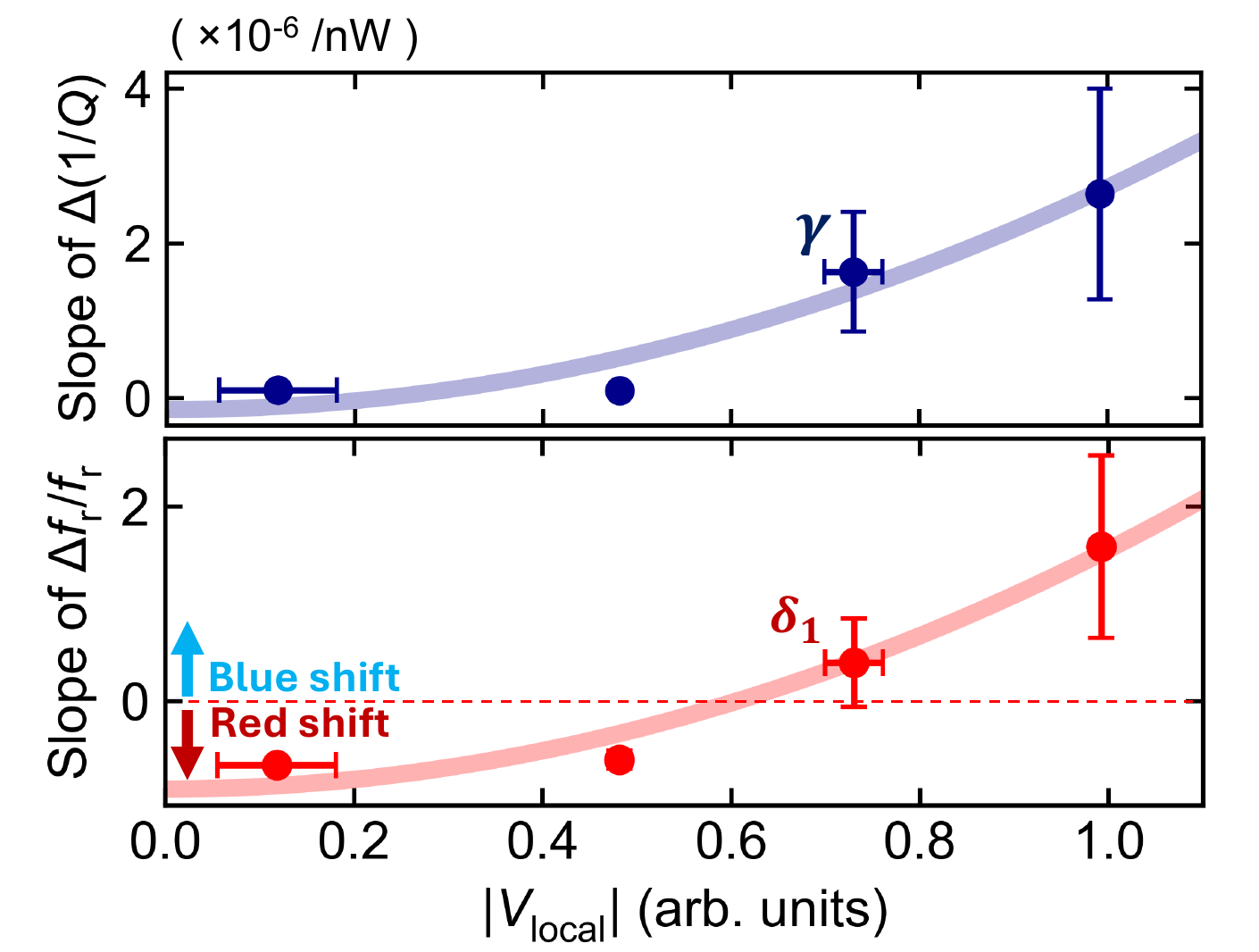}
	\caption{Relationship between local electric potential and changes in the resonance properties.
		(Top) Slope of $\Delta(1/Q)$ ($=\gamma$), and (Bottom) slope of $\Delta f_\mathrm{r}/f_\mathrm{r}$ ($=\delta_1$) as a function of the local potential, $|V_\mathrm{local}|$, in each mode and area. The average is calculated over data points with similar $|V_\mathrm{local}|$ values. The vertical and horizontal error bars correspond to the standard deviation of these data points. The fitting curves are the quadratic function ignoring the weight from the standard deviation.
	}
	\label{fig4}
\end{figure}

To analyze quantitatively  the observed spatial correlation between the shifts in ($\Delta(1/Q)$,  $\Delta f_\mathrm{r}/f_\mathrm{r}$) and the current density, the slopes of $\Delta(1/Q)$ ($\gamma$) and $\Delta f_\mathrm{r}/f_\mathrm{r}$ ($\delta_1$) are plotted as a function of $|V_\mathrm{local}|$ in Fig. \ref{fig4}. Here, the relative amplitude of the local potential, $|V_\mathrm{local}|$, is defined as $|V_\mathrm{local}|=|\cos(\sin^{-1}J(x,y))|$, where $J(x,y)$ is the normalized current-density distribution obtained from electromagnetic field simulations. 
Fig. \ref{fig4} shows that a higher local potential, which quantifies the local electric-field strength, results in larger slopes of $\Delta(1/Q)$ and $\Delta f_\mathrm{r}/f_\mathrm{r}$. Additionally, both $\gamma$ and $\delta_1$ exhibit quadratic dependence on $V_\mathrm{local}$.
This indicates that both $\Delta(1/Q)$ and $\Delta f_\mathrm{r}/f_\mathrm{r}$ are dominated by a mechanism other than quasiparticle generation, whose effects scale with the square of the resonator magnetic field (current density) \cite{PozarBook}.

\begin{figure}[t]
	\centering
	\includegraphics[width=8cm]{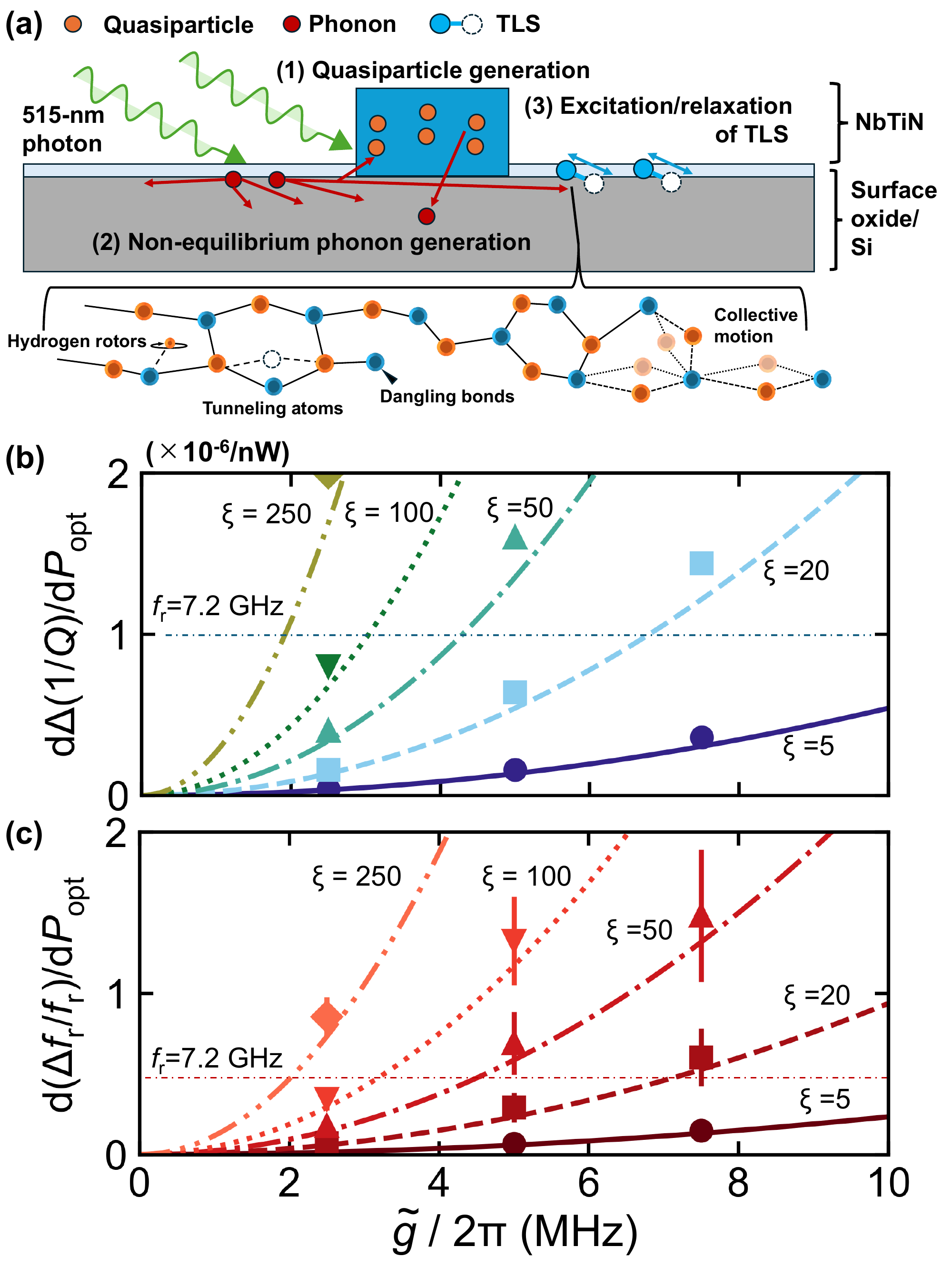}
	\caption{Schematic of the laser-induced effects and theoretical calculation based on the TLS model. (a) Laser illumination induces (1) quasiparticle generation in the superconductor, (2) nonequilibrium phonon generation, and (3) excitation/relaxation of TLSs via phonon absorption/emission.
	Some TLS formation mechanisms  in the amorphous material are shown \cite{Muller2019}. 
 (b) $\mathrm{d}\Delta(1/Q)/\mathrm{d}P_\mathrm{opt}$ and (c) $\mathrm{d} (\Delta f_\mathrm{r}/f_\mathrm{r})/\mathrm{d}P_\mathrm{opt}$ as a function of $g$ for different $\xi$. The lines are calculated using the analytical formula, while the dots with error bars are obtained from Monte Carlo simulations. The error bars correspond to the standard deviation over 100 trials. In each figure, the dotted line indicates the experimentally measured value at the voltage antinode of the third mode (7.2 GHz).
}
	\label{fig5}
\end{figure}

To explain the observed behaviors not accounted for by quasiparticle generation, we consider a microscopic model based on TLSs, which originate from defects, impurities, and other mechanisms \cite{Muller2019} at the metal and substrate surfaces. As TLS-induced loss depends on the local electric field \cite{Wang2020}, it is natural to consider the contribution  of TLSs.
Laser irradiation induces the following TLS-related processes (Fig. \ref{fig5}(a)). First, 515-nm photons incident on the superconductor generate quasiparticles, leading to the degradation of \(Q\) and a decrease in \(\Delta f_\mathrm{r}/f_\mathrm{r}\). Simultaneously, photons absorbed at the silicon surface generate nonequilibrium high-energy phonons via the recombination of excited electron--hole pairs. The recombination of quasiparticles in the superconductor also excites nonequilibrium phonons. Subsequently, these nonequilibrium phonons destroy Cooper pairs and alter the state of TLSs via phonon absorption and emission processes. The nonequilibrium population imbalance of TLSs   induces complex frequency shifts via the transverse and longitudinal coupling between the resonator and TLSs.

We quantitatively model the aforementioned process to reproduce the slopes of $\Delta(1/Q)$ and $\Delta f_\mathrm{r}/f_\mathrm{r}$
as follows. We assume that (1) the nonequilibrium phonon diffusion length scales as $\xi P_\mathrm{opt}$ thus the number of TLSs in the phonon-propagated volume  scales as $\hbar\omega_\mathrm{max}\rho_{\mathrm{TLS}}A\xi P_\mathrm{opt}$, where $\rho_\mathrm{TLS}$ is the TLS density of states per volume  (J$^{-1}$m$^{-3}$), $A$ is the cross sectional area around the nanowire for consideration, and $\omega_\mathrm{max}$ is the cutoff frequency for TLSs, (2) the state of TLSs inside $|x|<\xi P_\mathrm{opt}/2$ can be described as nonequilibrium parameters  $\tilde{S}$ ($\equiv\ev{\hat{\sigma}_z}$, $-1\leqq\ev{\hat{\sigma}_z}\leqq0$) and $\mathrm{d}\tilde{S}$  ($\equiv\mathrm{d}\ev{\hat{\sigma}_z}/\mathrm{d}\omega$), where $x$ is the distance from the laser spot, and (3) at high microwave driving powers, the contributions from TLSs originally resonant with the resonator without laser irradiation can be ignored. Then, considering the transverse ($\perp$) and longitudinal ($\parallel$) coupling between a TLS and a resonator and performing integration over related TLSs, the slopes of $\Delta(1/Q)$ and $(\Delta f_\mathrm{r}/f_\mathrm{r})$ are expressed as
\begin{equation}\label{eq:main_TLS_loss}
\dv{\Delta\qty(1/Q)}{P_\mathrm{opt}}=
2CK_{\parallel}\omega_\mathrm{r}\tilde{g}^2,
\end{equation}
\begin{equation}\label{eq:main_freq_shift}
\dv{\Delta f_\mathrm{r}/f_\mathrm{r}}{P_\mathrm{opt}}=C(K_\perp-\tilde{\Gamma}_1K_\parallel)\tilde{g}^2
,
\end{equation}
where $C=\hbar\rho_{\mathrm{TLS}}A\xi/\omega_\mathrm{r}$,  $K_{\parallel}=\tilde{\Gamma}_1\omega_\mathrm{max}\mathrm{d}\tilde{S}/(\tilde{\Gamma}_1^2+\omega_\mathrm{r}^2)$ is derived considering the longitudinal coupling, $K_\perp=(1+\tilde{S})\mathrm{ln}(\Delta_\mathrm{max}'/\Delta_{\mathrm{min}}')$ is the  coefficient for the transverse coupling, $\tilde{g}$ is the single-photon coupling between a resonator and TLSs averaged over related TLSs, $\tilde{\Gamma}_1$ is the averaged relaxation time of TLSs, and $\Delta_\mathrm{max}'=\omega_\mathrm{max}\ (\Delta_\mathrm{min}'=\omega_\mathrm{r})$ is the cutoff detuning used for the frequency-shift calculation (see Supplementary Discussion for details of the equations including derivations). 

Figures \ref{fig5} (b)(c) plot $\mathrm{d}\Delta(1/Q)/\mathrm{d}P_\mathrm{opt}$ and $\mathrm{d}(\Delta f_\mathrm{r}/f_\mathrm{r})/\mathrm{d}P_\mathrm{opt}$ as functions of $\tilde{g}$ and $\xi$ for fixed parameters, $\tilde{S}=0$ (complete mixed state), $2\pi\mathrm{d}\tilde{S}=1/400$ MHz$^{-1}$, $\rho_\mathrm{TLS} = 1\times10^{45}$ J$^{-1}$m$^{-3}$ \cite{Constantin2007,Skacel2014,Burnett2016,Lisenfeld2019}, $A=tw$ (TLS layer thickness $t = 2$ nm; effective width for consideration $w = 500$ nm), $\omega_\mathrm{r}/2\pi = 7$ GHz, $\tilde{\Gamma}_1/2\pi$ = 16 MHz,   $\Delta_\mathrm{max}'/2\pi\simeq\omega_\mathrm{max}/2\pi=1000$ GHz, and  $\Delta_\mathrm{min}'=\omega_\mathrm{r}$. The model can reproduce the experimentally observed quadratic dependence of $\mathrm{d}\Delta(1/Q)/\mathrm{d}P_\mathrm{opt}$ and $\mathrm{d}(\Delta f_\mathrm{r}/f_\mathrm{r})/\mathrm{d}P_\mathrm{opt}$ on $|V_\mathrm{local}|$ because $\tilde{g}\propto|V_\mathrm{local}|$. For the chosen parameters with $\tilde{g}/2\pi\sim$ 2--8 MHz and $\xi$ = 20--250 m/W, we obtain values close to the observed slopes for both $\Delta(1/Q)$ and $\Delta f_\mathrm{r}/f_\mathrm{r}$, $\sim1\times10^{-6}$ /nW and $\sim0.5\times10^{-6}$ /nW for $f_\mathrm{r}=7.2$ GHz at the voltage antinode. Regarding parameter validity , a $g$ of several MHz is observed in the high-$L_\mathrm{k}$ resonator \cite{Kristen2023}. A value of $\xi$ of $20$--$250$ m/W indicates that the effective phonon diffusion length is $\pm$10--125 $\mu$m at 1 $\mu$W laser power. The effective phonon  diffusion length of $\mathcal{O}(10)$ $\mu$m/$\mu$W can explain that the spatial resolution of our system is limited to $\sim10$ $\mu$m (see Supplementary Methods). If $\xi\sim\mathcal{O}(1000)$, the spatial dependent response would be averaged out since the length of one side of the square-shaped resonator is $\sim$ 400 $\mu$m.
A Monte Carlo simulation that assumes  random TLS positional and frequency distributions also reproduces similar slopes (dots with error bars in Figs. \ref{fig5} (b)(c)) (see Supplementary Discussion for the detailed theoretical model).

Furthermore, we must consider (1) the frequency dependence of $\mathrm{d}\Delta(1/Q)/\mathrm{d}P_\mathrm{opt}$ and $\mathrm{d}(\Delta f_\mathrm{r}/f_\mathrm{r})/\mathrm{d}P_\mathrm{opt}$, (2) the spatial-mode distribution, and (3) the inhomogeneous TLS distribution, to understand the observed behaviors remaining unexplained: smaller  $\mathrm{d}\Delta(1/Q)/\mathrm{d}P_\mathrm{opt}$ and $\mathrm{d}(\Delta f_\mathrm{r}/f_\mathrm{r})/\mathrm{d}P_\mathrm{opt}$ at higher modes, the deviation from the global trend at Area 4, and large standard deviation in Fig. \ref{fig4}. Based on Eqs. \eqref{eq:main_TLS_loss}\eqref{eq:main_freq_shift}, $\mathrm{d}\Delta(1/Q)/\mathrm{d}P_\mathrm{opt}\propto1/\omega_\mathrm{r}$ and  $\mathrm{d}(\Delta f_\mathrm{r}/f_\mathrm{r})/\mathrm{d}P_\mathrm{opt}\propto \text{const.}$ as a function of the  resonance frequency. In addition, at higher  modes with rapid spatial variation, the local response can be averaged out. For instance, since the distance between nodes and antinodes for the fifth mode is $\sim$150 $\mu$m, the spatial variation in $V_\mathrm{local}$ is smoothed out assuming a large $\xi>100$ $\mu$m/$\mu$W. These explain that higher the resonance mode, smaller the $\mathrm{d}\Delta(1/Q)/\mathrm{d}P_\mathrm{opt}$ and $\mathrm{d}(\Delta f_\mathrm{r}/f_\mathrm{r})/\mathrm{d}P_\mathrm{opt}$ (observation of the subtle shift instead of the blue shift). Additionally, as the TLS position and frequency distributions are random, the resonator response exhibits a large standard deviation that depends on the number and frequencies of nearby TLSs. This effect is remarkable for $\Delta f_\mathrm{r}/f_\mathrm{r}$ in Monte Carlo simulations. Thus, the deviation from the global trend in the third and fifth modes at Area 4 can be explained by considering that the number of relevant TLSs is larger at frequencies between these modes. These factors, (1)(2)(3), cause the variation in optical responses that are dependent on resonant modes and positions in addition to $|V_\mathrm{local}|$, resulting in the large standard deviation in Fig. \ref{fig4}, which shows averaged values only focusing on $|V_\mathrm{local}|$.

Finally, we note the influence of equilibrium temperature shifts as discussed in Refs.  \cite{Wang2013a, Janssen2014}. Throughout the discussion above, we ignore the temperature-dependent shift of TLS permittivity because the experimentally estimated temperature-induced shifts are too small to account for the laser-induced shifts in resonance properties (see Supplementary Discussion for details). The point is that laser-induced high-energy nonequilibrium phonons excite high-energy TLSs ($\hbar\omega_\mathrm{TLS}>\hbar\omega_\mathrm{r},\ k_BT$) that are never excited under equilibrium conditions. This is why the blue shift and large degradation in $Q$ are only observed in optical-illumination experiments, not for temperature dependence measurements.

In conclusion, we have measured the spatial and mode-dependent optical response of the superconducting nanowire microwave resonator. This method, based on laser-scanning microscopy, enables precise control of the irradiation position. We have demonstrated that the local optical response strongly depends on the local electric potential of the resonator, which can be explained by interactions between TLSs and nonequilibrium phonons. These insights provide guidance for the design and operation of devices that couple microwave and optical fields, including quantum transducers and superconducting detectors. Moreover, the developed local optical-illumination technique can be applied  to superconducting qubits, enabling the investigation of their critical behavior in quantum computing chips dominated by the interplay of high-energy particle irradiation, nonequilibrium phonons, TLSs, and the coherence of superconducting circuits.

\section*{Acknowledgements}
\section*{Funding}
H. Kosaka acknowledges the funding support from the Japan Science and Technology Agency (JST) Moonshot R$\&$D grant (JPMJMS2062) and a JST CREST grant (JPMJCR1773). H. Kosaka also acknowledges the Ministry of Internal Affairs and Communications (MIC) for funding the research and development to construct a global quantum cryptography network  (JPMI00316), and the Japan Society for the Promotion of Science (JSPS) Grants-in-Aid for Scientific Research (20H05661, 20K20441).
\section*{Authorship contribution}
R. H. performed the experiments. H. Kurokawa designed the experiments and wrote the manuscript. K. T. and H. T. fabricated the sample. H. Kosaka  supervised the project.

\section*{Competing interests}
The authors declare no competing interests.

\section*{Data availability}
All data required for evaluating the conclusions in the paper  are present in the paper and the repository (DOI 10.5281/zenodo.15680751). 

\clearpage
\input{mr_opt_response_change_model_12_arxiv.bbl}

\clearpage          
\onecolumngrid      

\beginsupplement    
\begin{center}
	\section*{Supplementary Materials for\\ Spatially Resolved Optical Responses of \\a  Superconducting Nanowire Microwave Resonator}
 \normalsize
Rento Hirotsuru$^{1}$,
Hodaka Kurokawa$^{2\ast}$,
Kazuyo Takaki$^{3}$,
Hirotaka Terai$^{2,3}$,
and Hideo Kosaka$^{1,2\ast}$\par
\vspace{6pt}\itshape
$^{1}$Department of Physics, Graduate School of Engineering Science,\\
Yokohama National University, 79-5 Tokiwadai, Hodogaya, Yokohama, 240-8501, Japan\\
$^{2}$Quantum Information Research Center, Institute of Advanced Sciences,\\
Yokohama National University, 79-5 Tokiwadai, Hodogaya, Yokohama, 240-8501, Japan\\
$^{3}$National Institute of Information and Communications Technology, \\
588-2, Iwaoka, Nishi-ku, Kobe, Hyogo 651-2492, Japan\\[6pt]
\normalsize
\end{center}

\input{spatial_res_high_Lk_supply_03.tex}


\end{document}

%% file: mr_opt_response_change_model_12_arxiv.bbl
%

%% file: spatial_res_high_Lk_supply_03.tex
	\title{Supplementary Information for: Spatially Resolved Optical Responses of a Superconducting Nanowire Microwave Resonator}
	
	
\author{Rento Hirotsuru}
\affiliation{Department of Physics, Graduate School of Engineering Science,
	Yokohama National University, Yokohama 240-8501, Japan}

\author{Hodaka Kurokawa}
\email{kurokawa-hodaka-hm@ynu.ac.jp}            
\affiliation{Quantum Information Research Center, Institute of Advanced Sciences,
	Yokohama National University, Yokohama 240-8501, Japan}

\author{Kazuyo Takaki}
\affiliation{National Institute of Information and Communications Technology,
	Kobe 651-2492, Japan}

\author{Hirotaka Terai}
\affiliation{Quantum Information Research Center, Institute of Advanced Sciences,
	Yokohama National University, Yokohama 240-8501, Japan}
\affiliation{National Institute of Information and Communications Technology,
	Kobe 651-2492, Japan}

\author{Hideo Kosaka}
\email{kosaka-hideo-yp@ynu.ac.jp}               
\affiliation{Department of Physics, Graduate School of Engineering Science,
	Yokohama National University, Yokohama 240-8501, Japan}
\affiliation{Quantum Information Research Center, Institute of Advanced Sciences,
	Yokohama National University, Yokohama 240-8501, Japan}

\maketitle
	
\RestoreTOC            
\tableofcontents       

\section{Supplementary Methods}
\subsection*{Sample fabrication}
The fabrication process of the NbTiN nanowire microwave resonator is as follows. First, a NbTiN thin film of approximately 10-nm thickness is deposited on a hydrogen-terminated silicon wafer by magnetron sputtering, where the silicon wafer has a resistivity of 20 k$\Omega$cm or higher. Patterning of the wafer is performed using an EB lithography system (ELIONIX ELS-125) for the nanowire and photolithography with a maskless aligner (Heiderberg MAL150) for the large area such as bonding pads. After forming alignment marks for the subsequent EB lithography and photolithography, aligned to these marks, the nanowire with a line width of 120 nm is patterned by EB lithography using a 150-nm-thick positive EB resist ZEP530A. Next, a 2-nm-thick MgO is deposited as a mask layer for etching the NbTiN thin film using CF$_4$ gas, and the ZEP530A is lifted-off using N-Methyl-2-Pyrrolidone (NMP). Then, a large-area pattern is patterned on the nanowire using photolithography, and the NbTiN thin film is etched by RIE using CF$_4$ gas. After removing the photoresist using NMP, wet etching using buffered HF is performed to remove the MgO mask layer remaining on the nanowire. The line width of the fabricated nanowire is checked by SEM observation. It is found to be approximately 150 nm for the design value of 120 nm. The resultant nanowire resonator comprises a loop structure with a 5-\SI{}{\micro\metre} gap fabricated from a 10-nm-thick, 150-nm-wide, and 1.5-mm-long NbTiN nanowire (see Methods for details on fabrication). Two \SI{5}{\micro\metre}$\cross$\SI{5}{\micro\metre} pad structures are incorporated at the gap for optical alignment.

\subsection*{Experimental setup}
\begin{figure}[b]
	\centering
	\includegraphics[width=12cm,clip]{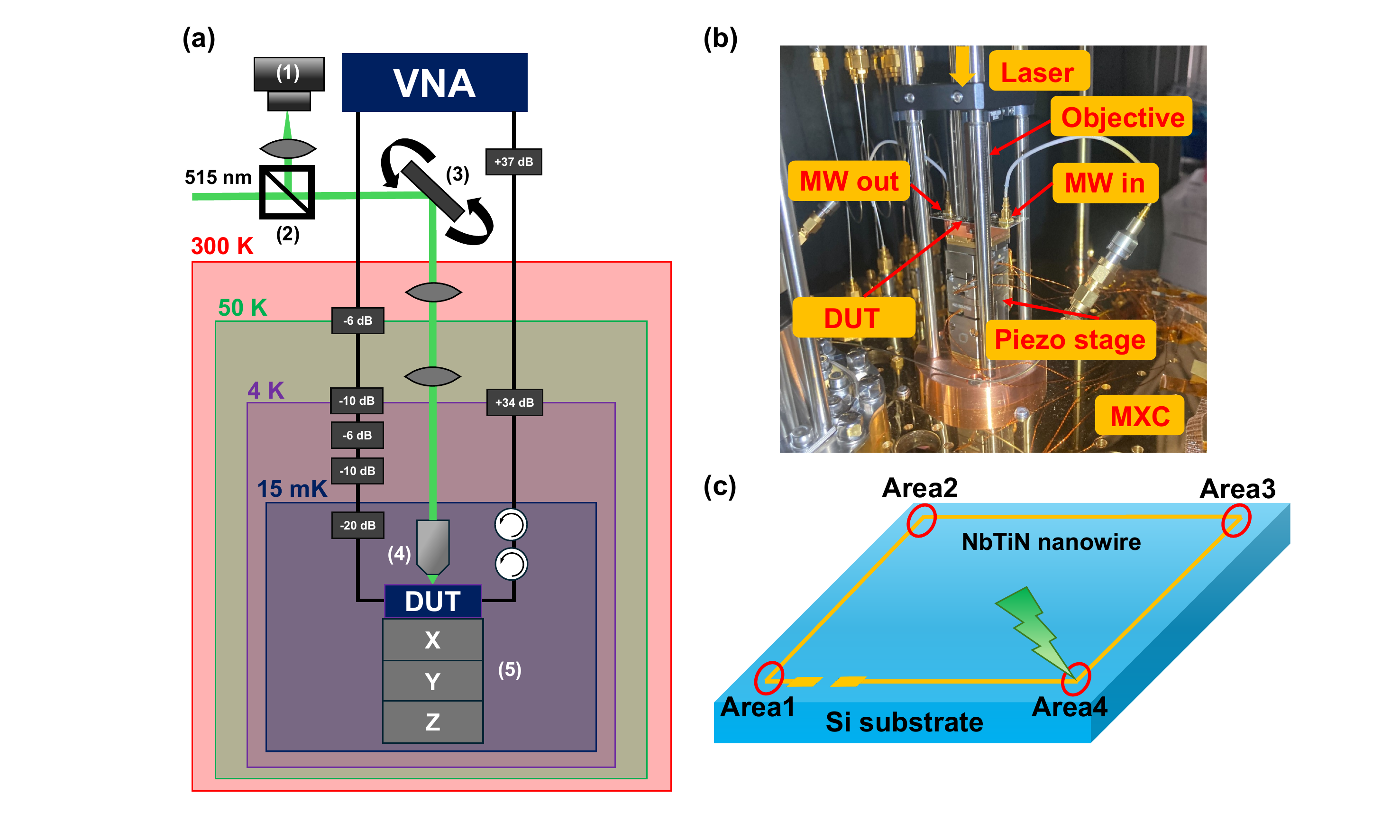}
	\caption{Experimental setup. (a) Schematic of the optical and microwave system integrated with the dilution refrigerator. (1) CCD camera, (2) beamsplitter, (3) galvanometer mirror, (4) objective lens, (5) 3-axis piezo-positioner. (b) Photograph of the experimental system on the mixing chamber plate (MXC). (c) Schematic of the microwave resonator used in the experiment. The laser is focused on the corners of the resonator (Area 1-4).
	}
	\label{fig:setup}
\end{figure}

All experiments are performed in a closed-cycle dilution refrigerator (LD-400, Bluefors) in the presence of an ambient magnetic field (Fig. \ref{fig:setup}). To reduce blackbody radiation from the high-temperature plates, a copper plate with a 10-mm-diameter hole is mounted on each plate. The mixing chamber plate is maintained at approximately 15 mK, as monitored by a thermometer attached to it. A copper sample holder is mounted on the XYZ piezoelectric nanopositioners (ANPx101/RES/LT$\times2$, ANPz101/RES/LT, Attocube) via a thermal link. These positioners are used for coarse alignment of the sample with a CCD camera. The sample's surface is identified as the position where the reflected spot image on the CCD camera is minimized. Relatively large structures, such as the waveguide, can be distinguished in the CCD image, whereas the nanowire cannot be resolved.

The optical system is configured as follows. A 515-nm laser (Cobolt, Hubner Photonics) is used for the laser irradiation experiments. The laser is scanned over the sample using a two-axis galvanometer mirror (GVS202, Thorlabs). Two lenses are inserted between the galvanometer mirror and an objective lens (LT-APO/VISIR/0.82, Attocube) to form a 4-$f$ system, which maximizes the microscope's field of view. The laser, focused by the objective lens, is directed onto the surface of the sample. Since laser irradiation around the resonator modulates the microwave transmission spectrum, changes in the spectrum allow us to roughly estimate the focusing position while moving the piezo-positioner. After roughly locating the corners of the resonator, we acquire a laser-scanning microwave transmission spectrum map to determine the precise position for subsequent laser-irradiation experiments.

The microwave setup is configured as follows. A vector network analyzer (VNA) (P9373A, Keysight) is used to measure the transmission spectrum ($S_{21}$) of the resonator. The probe microwave is attenuated by a total of 52 dB. The input power is calculated considering the output power from VNA and the 52-dB attenuation, and the finite cable loss is not calibrated. The probe signal transmitted through the sample is then amplified by a high-electron mobility-transistor (HEMT) amplifier and a room-temperature amplifier before being detected by the VNA. Here, we insert two circulators between the HEMT and the resonator.

\subsection*{Spatial Resolution of the Experimental System}

\begin{figure}[bt]
	\centering
	\includegraphics[width=14cm,clip]{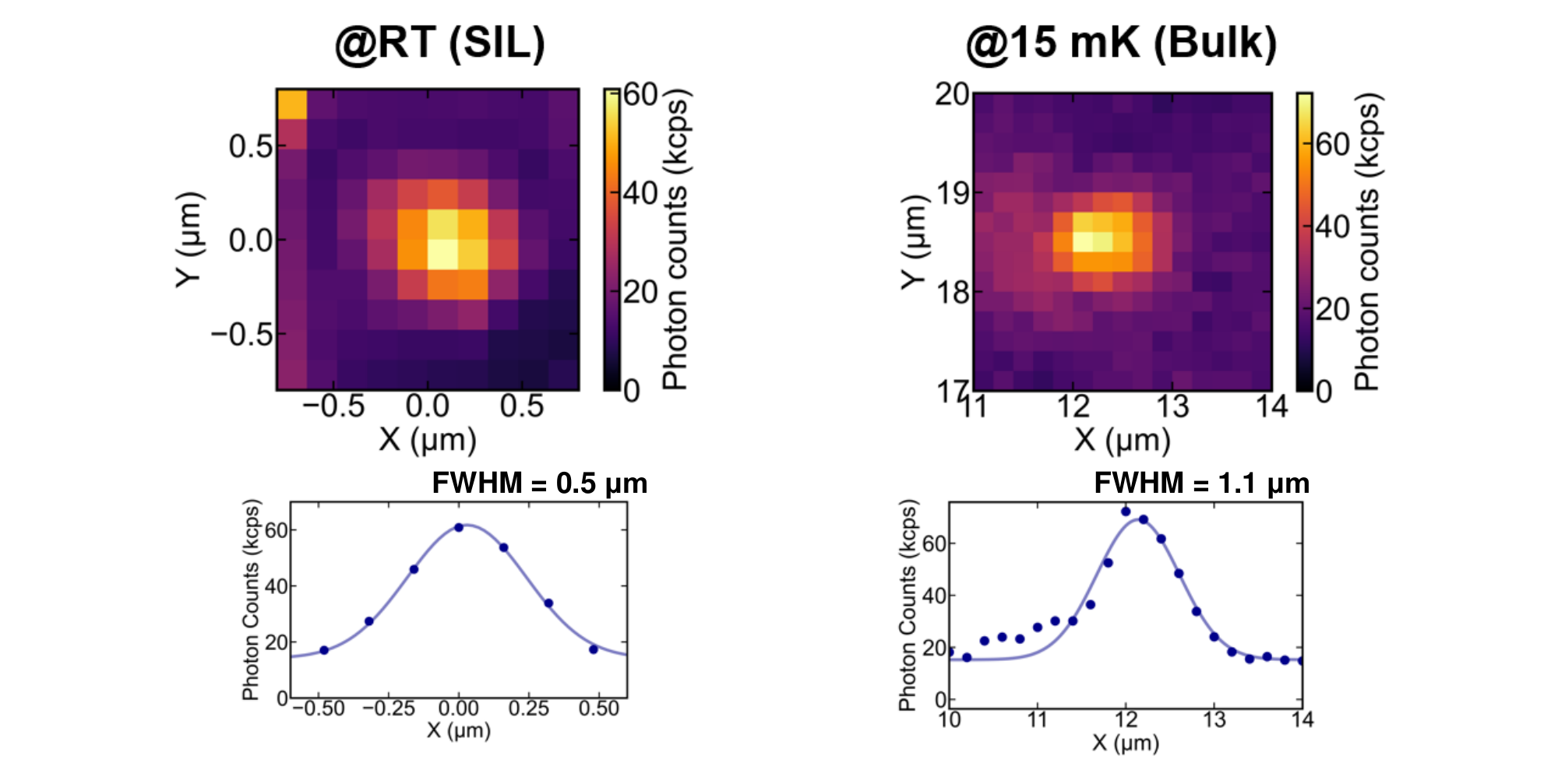}
	\caption{Photoluminescence (PL) maps of a nitrogen-vacancy center in diamond. (Left) Room temperature measurement with a full-width half-maximum (FWHM) of 0.5 \(\mu\)m; (Right) low-temperature measurement at 15 mK with a FWHM of 1.1 \(\mu\)m. Different diamond samples were used: a nitrogen-vacancy center inside a solid immersion lens (SIL) for the room temperature experiment, and a nitrogen-vacancy center in bulk diamond for the low-temperature experiment.}
	\label{fig:resolution}
\end{figure}

In this section, we discuss the spatial resolution of our microscopy system. Figure~\ref{fig:resolution} shows a photoluminescence (PL) map of a nitrogen-vacancy center in diamond. The left map, acquired at room temperature, exhibits a full-width half-maximum (FWHM) of 0.5 \(\mu\)m. Given that the beam waist for a 515-nm laser is approximately \(r \times 0.6\) \(\mu\)m (assuming \(NA=0.82\) and \(r=2/\pi\) for the full-width \(1/e^2\) and \(r=1.18/\pi\) for the full-width half maximum), this FWHM is close to the diffraction-limited value. In contrast, a low-temperature measurement, performed on a different diamond sample, shows a FWHM of 1.1 \(\mu\)m, possibly due to vibrations of the dilution refrigerator and variations in excitation power between measurements. These results indicate that our microscopy system can focus the 515-nm laser nearly at the diffraction limit, ruling out the possibility that the laser beam waist is broader than expected.

\begin{figure}
	\centering
	\includegraphics[width=14cm,clip]{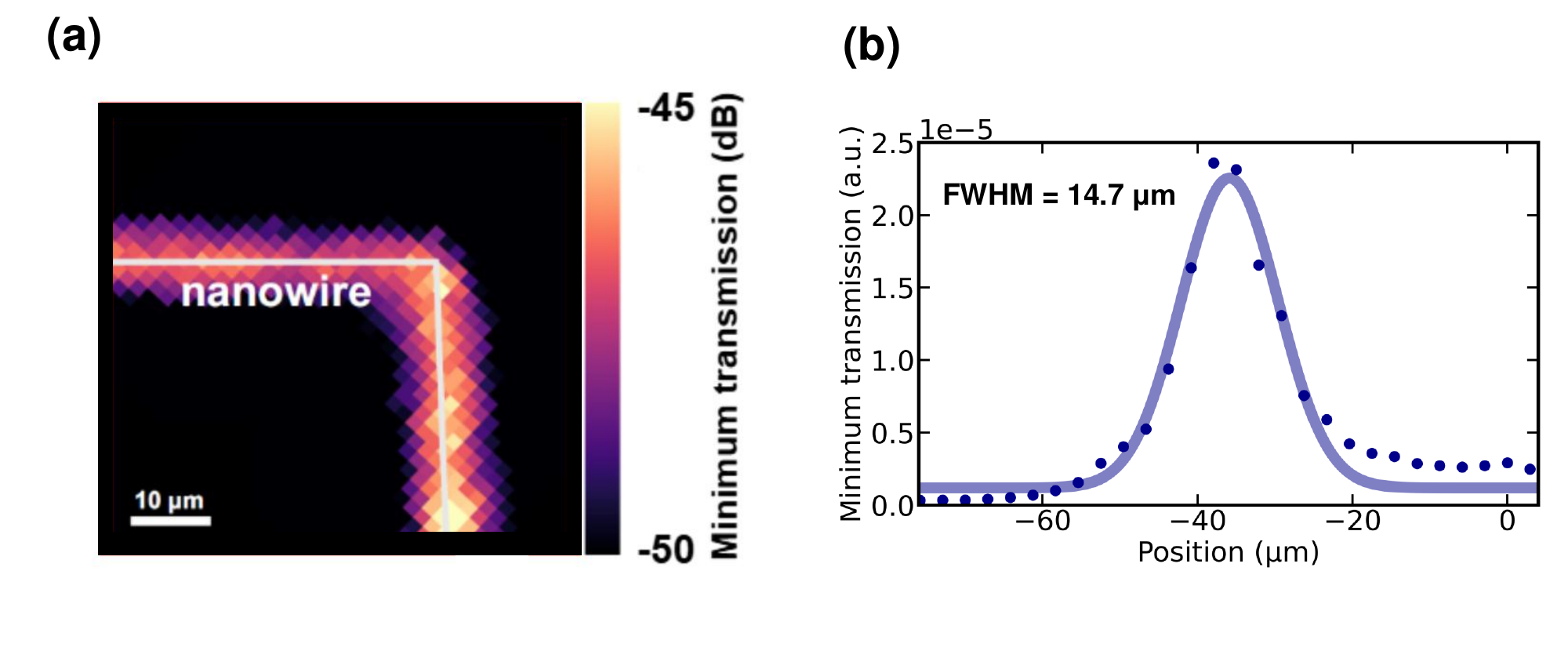}
	\caption{(a) Laser-scanning map at Area 3. (b) Cross section of (a), indicating FWHM is 14.7 $\mu$m. }
	\label{fig:nanowire_resolution}
\end{figure}

Figure~\ref{fig:nanowire_resolution}(a) shows the result of the laser-scanning microwave spectroscopy at Area 3, while Fig. \ref{fig:nanowire_resolution}(b) displays its one-dimensional profile, revealing an FWHM of 14.7 \(\mu\)m. In contrast to the spatial resolution expected from the beam waist, a rather broad optical response is observed. The origin of the optical response in broader area than the beam waist may be attributed to the diffusion of laser-induced phonons which is discussed in the following.

\subsection*{Simulation of the Resonance Mode and the Current Distribution}

\begin{figure}[b]
	\centering
	\includegraphics[width=16cm,clip]{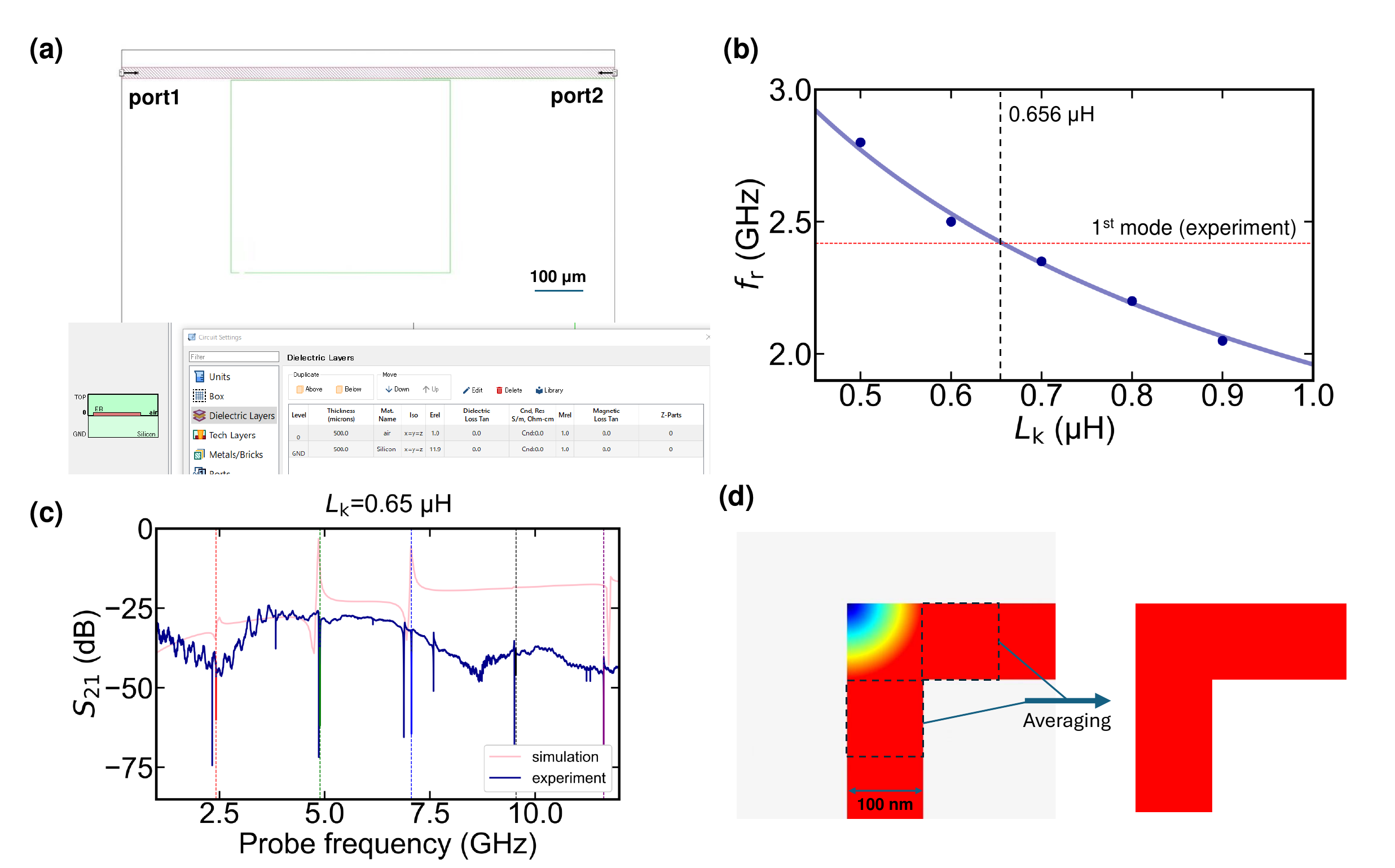}
	\caption{Electromagnetic field simulation.  (a) Resonator shape used in the simulation. The red waveguide is a perfect conductor, and the green nanowire is a superconductor with a sheet inductance due to the kinetic inductance. The simulation environment is also shown. (b) Simulated resonance frequencies of the first mode as a function of the total kinetic inductance of the nanowire. The data are fitted assuming \(f_\mathrm{r}\propto1/\sqrt{L_\mathrm{k}}\). (c) Simulated transmission spectrum of the resonator for \(L_\mathrm{k} = 0.65\,\mu\mathrm{H}\) (red line) compared with the experimental transmission spectrum (blue line). (d) Treatment of current crowding at the corner. The normalized current density, averaged over a \(200\times100\) nm\(^2\) area, is used for the analysis in the main text.
	}
	\label{fig:simulation}
\end{figure} 

We use the electromagnetic field analysis software (Level3Gold Antenna, Sonnet) to simulate the current distribution in the resonator. Figure~\ref{fig:simulation}(a) shows the configuration and settings used for the simulation. To simplify the calculation, the width of the nanowire is set to 100 nm. To estimate the kinetic inductance of the resonator, the resonance frequencies of the first mode are plotted as a function of the kinetic inductance, \(L_\mathrm{k}\) (Fig. \ref{fig:simulation}(b)). A value of 0.656 \(\mu\)H reproduces the experimentally observed frequency of the first resonance mode. Figure~\ref{fig:simulation}(c) shows both the simulated transmission spectrum of the resonator for \(L_\mathrm{k} = 0.65\   \mu\)H and the experimental transmission spectrum. A slight deviation is observed between the resonance frequencies obtained from the experiment and simulation at higher resonance modes, the origin of which is unknown. Additionally, note that the current density used in the main text analysis represents the spatial average over an area of approximately 200 × 100 nm\(^2\) adjacent to the corner to avoid the effects of current crowding in the simulation (Fig. \ref{fig:simulation}(d)). Since we do not observe an anomalous behavior at the corners throughout the laser-scanning experiment, we consider that this treatment can be justified. For a more refined experiment, rounding the corner may be useful to make the current density uniform.

\subsection*{Resonator Characterization}

The transmission spectrum, \(S_{21}\), of a hanger-type resonator employed in the main text is typically expressed as \cite{Wang2020}:
\begin{equation}\label{eq:int_3dBwidth}
	S_{21}(f)=1-\frac{Q_\mathrm{tot}/Q_\mathrm{ext}}{1+2iQ_\mathrm{tot}(f-f_\mathrm{r})/f_\mathrm{r}},
\end{equation}
where $Q_\mathrm{tot}$ (\(1/Q_\mathrm{tot} = 1/Q_\mathrm{int} + 1/Q_\mathrm{ext}\)) is the total quality factor, with \(Q_\mathrm{int}\) and \(Q_\mathrm{ext}\) representing the internal and external quality factors, respectively. More sophisticated treatments, which include impedance mismatch and non-ideal couplings, are introduced and summarized in Refs. \cite{Gao2008,Wang2020,Chen2022}. It should be noted that the quality factor in the main text corresponds to the internal quality facotr. The subscript, int, is omitted for simplicity.

The half-power bandwidth, \(2\delta f'\) (with \(\delta f' = |f' - f_\mathrm{r}|\) defined by the condition \(2|S_{21}(f_\mathrm{r})|^2 = |S_{21}(f')|^2\)), measured from the bottom of the dip, can be expressed as
\begin{equation}\label{eq:int_loss_bandwidth}
	2\pi \times 2\delta f' = \frac{\kappa_\mathrm{int}}{\sqrt{1-\frac{2\kappa_\mathrm{int}^2}{\kappa_\mathrm{tot}^2}}},
\end{equation}
where \(\kappa_\mathrm{int} = \omega_\mathrm{r}/Q_\mathrm{int}\) is the internal decay rate, \(\kappa_\mathrm{ext} = \omega_\mathrm{r}/Q_\mathrm{ext}\) is the external decay rate, and \(\kappa_\mathrm{tot} = \kappa_\mathrm{int} + \kappa_\mathrm{ext}\) is the total decay rate of the resonator. For an overcoupled resonator, such that \(\kappa_\mathrm{ext} \gg \kappa_\mathrm{int}\), Eq.~\eqref{eq:int_loss_bandwidth} reduces to \(\kappa_\mathrm{int}\), indicating that the 3-dB bandwidth from the dip corresponds to the internal loss rate. 

On the other hand, another half-power bandwidth, \(2\delta f'' = 2\,|f'' - f_\mathrm{r}|\) (with \(0.5 = |S_{21}(f'')|^2\)), can be related to the external loss as follows:
\begin{equation}
	2\pi \times2\delta f'' = \kappa_\mathrm{ext}\sqrt{1+\frac{2\kappa_\mathrm{int}}{\kappa_\mathrm{ext}}}.
\end{equation}
For an overcoupled resonator, this equation reduces to \(\kappa_\mathrm{ext}\), indicating that \(2\delta f''\) corresponds to the external loss of the resonator.

The number of photons in the resonator is calculated as follows based on the procedure used in Ref \cite{Bruno2015}. The power dissipated inside the resonator, $P_\mathrm{loss}$, is
\begin{equation}
	P_\mathrm{loss} = P_\mathrm{in} - P_\mathrm{refl} -P_\mathrm{trans},
\end{equation}
where $P_\mathrm{in}$ is the input microwave power, $P_\mathrm{refl} = |S_{11}|^2 P_\mathrm{in}$ is the reflected power, $P_\mathrm{trans} = |S_{21}|^2 P_\mathrm{in}$ is the transmitted power. According to Ref. \cite{Chen2022}, for a hanger-type resonator,
\begin{equation}
\begin{aligned}
	S_{11} &= -\frac{Q_\mathrm{tot}/Q_\mathrm{ext}}{1+2iQ_\mathrm{tot}(f-f_\mathrm{r})/f_\mathrm{r}},\  |S_{11}|^2= \frac{(Q_\mathrm{tot}/Q_\mathrm{ext})^2}{1+4Q_\mathrm{tot}^2(f-f_\mathrm{r})^2/f_\mathrm{r}^2},
	\\
	S_{21}&=1-\frac{Q_\mathrm{tot}/Q_\mathrm{ext}}{1+2iQ_\mathrm{tot}(f-f_\mathrm{r})/f_\mathrm{r}},
	\ 
	|S_{21}|^2 = 1 - \frac{2Q_\mathrm{tot}/Q_\mathrm{ext}-(Q_\mathrm{tot}/Q_\mathrm{ext})^2}{1+4Q_\mathrm{tot}^2(f-f_\mathrm{r})^2/f_\mathrm{r}^2}.
\end{aligned}
\end{equation}
Thus, 
\begin{equation}
	P_\mathrm{loss} = \frac{2Q_\mathrm{tot}/Q_\mathrm{ext}(1-Q_\mathrm{tot}/Q_\mathrm{ext})}{1+4Q_\mathrm{tot}^2(f-f_\mathrm{r})^2/f_\mathrm{r}^2}P_\mathrm{in}
	=\frac{2\kappa_\mathrm{i}\kappa_\mathrm{ext}}{\kappa_\mathrm{tot}^2+4(\omega-\omega_\mathrm{r})^2}P_\mathrm{in}.
\end{equation}
Assuming that 
\begin{equation}
	P_\mathrm{loss} = \ev{n_\mathrm{cav}}\times \hbar\omega_\mathrm{r} \times \kappa_\mathrm{int},
\end{equation}
the number of photons in the resonator, $n_\mathrm{cav}$, can be expressed as
\begin{equation}
\begin{aligned}
n_\mathrm{cav}&=
\frac{2\kappa_\mathrm{ext}}{\kappa_\mathrm{tot}^2+4(\omega-\omega_\mathrm{r})^2}
\frac{P_\mathrm{in}}{\hbar\omega_\mathrm{r}}.
\end{aligned}
\end{equation}
For $\omega=\omega_\mathrm{r}$, this equation reduces to 
\begin{equation}
	\mathrm{cav}=
	\frac{2\kappa_\mathrm{ext}}{\kappa_\mathrm{tot}^2}
	\frac{P_\mathrm{in}}{\hbar\omega_\mathrm{r}}.
\end{equation}

In some Refs. \cite{Burnett2018a,Wang2020}, a factor considering the impedance mismatch between the transmission line and the resonator, $Z_0/Z_\mathrm{r}$, is multiplied to the expression above. However, since the above derivation does not explicitly include the impedance, we do not consider impedance difference explicitly in the following estimation of $n_\mathrm{cav}$.

\begin{figure}[tb]
	\centering
	\includegraphics[width=13cm,clip]{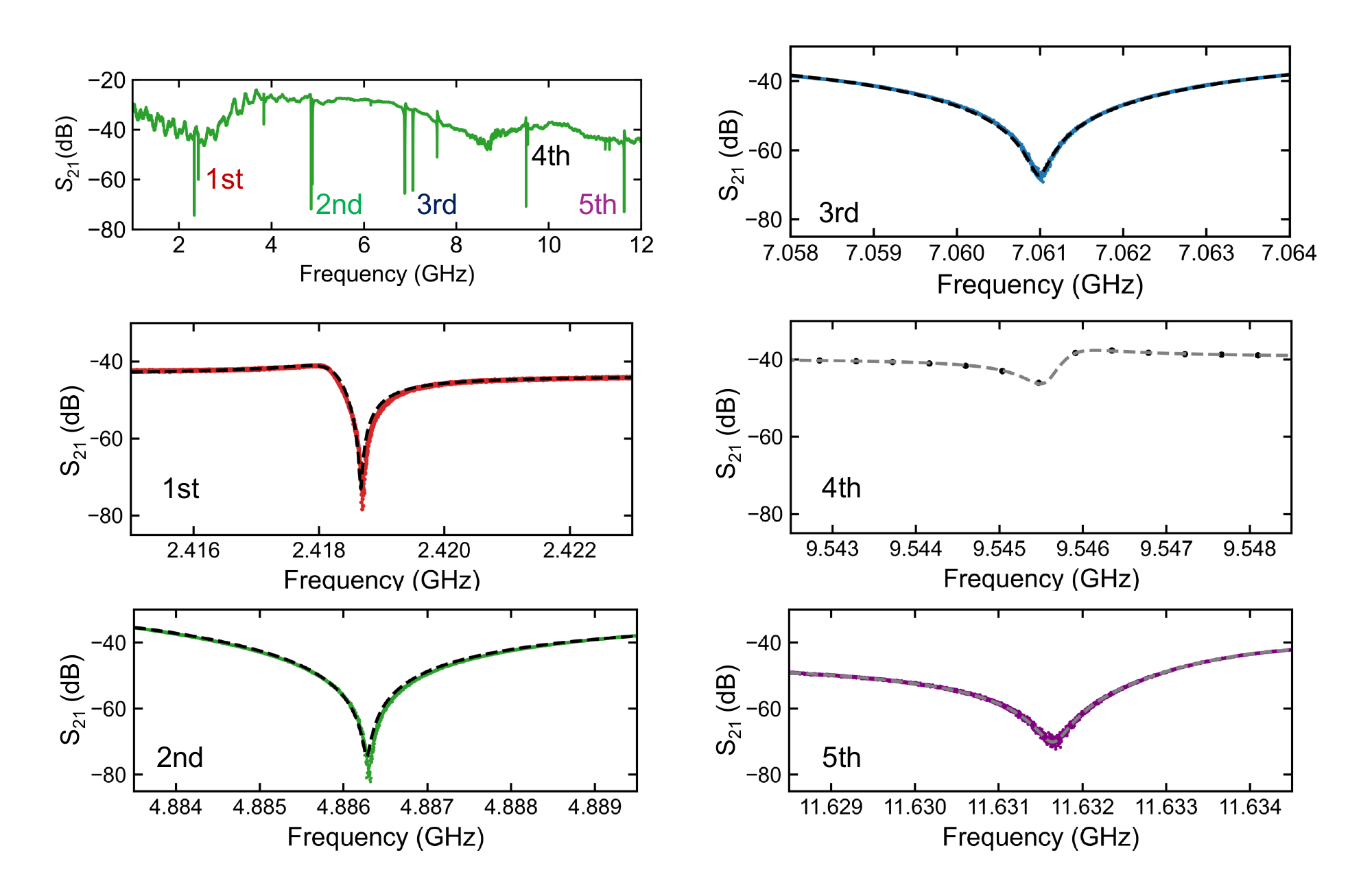}
	\caption{Resonator's spectrum from the first to the fifth mode. The dotted lines are the fitted curves.
	}
	\label{fig:spectrum}
\end{figure} 

Figure \ref{fig:spectrum} shows the resonance spectrum of the first to fifth modes. In Table \ref{tab:sammary_para}, we summarize the resonance frequencies, the external quality factor, the internal quality factor, and the number of photons for each mode based on the equations introduced above. For the fit of the transmission signal from the actual device, we used the function including the circuit asymmetry and a finite loss and phase delay in the transmission line as \cite{Chen2022}
\begin{equation}\label{eq:real_fit_S21}
\begin{aligned}
S_{21}(f)&=Ae^{-i(2\pi f\tau+\alpha)}\left(1-\frac{e^{i\phi}Q_\mathrm{tot}/Q_\mathrm{ext}}{1+2iQ_\mathrm{tot}(f-f_\mathrm{r})/f_\mathrm{r}}\right)
\\
&=Ae^{-i(2\pi f\tau+\alpha)}\left(1-\frac{Q_\mathrm{tot}/(Q_\mathrm{ext,real}+iQ_\mathrm{ext,imag})}{1+2iQ_\mathrm{tot}(f-f_\mathrm{r})/f_\mathrm{r}}\right)
,
\end{aligned}
\end{equation}
where $Ae^{-i(\omega\tau+\alpha)}$ corresponds to the damping and phase shifts in the transmission line, $e^{i\phi}$ originates from the circuit asymmetry. According to Ref. \cite{Chen2022}, the external quality factor is defined to be  $1/\mathrm{Re}(1/(Q_\mathrm{ext,real}+iQ_\mathrm{ext,imag}))$ and  the internal quality factor can be expressed as $1/Q_\mathrm{int}=1/Q_\mathrm{tot}-\mathrm{Re}(1/(Q_\mathrm{ext,real}+iQ_\mathrm{ext,imag}))$. It can be seen that the resonator used in the experiment is in the overcoupled regime such that $Q_\mathrm{int}>Q_\mathrm{ext}$. Thus, for the analysis of the experimental data in the main text, we fitted only the region around the resonance dip using the Lorentzian function, thereby obtaining a 3-dB bandwidth corresponding to the internal quality factor. Although 15--20$\%$ deviations in the absolute value of $Q_\mathrm{int}$ are observed at maximum between the fitting using Eq. \eqref{eq:real_fit_S21} and the fitting using the Lorentzian function, the Lorentzian fit is enough for capturing the direction of the shifts in $Q_\mathrm{int}$ and $f_\mathrm{r}$ as shown in the main text.

\begin{table}[]
	\centering
	\caption{Summary of parameters of the resonator used for the experiments}
	\label{tab:sammary_para}
	\begin{tabular}{l|llll}
		\hline
		& 1st mode         & 2nd mode                             & 3rd mode                             & 5th mode                             \\ \hline
		Frequency (GHz)                  & 2.418            & 4.884                                & 7.061                                & 11.63                                \\
		$Q_\mathrm{int}$ (Eq.\eqref{eq:real_fit_S21})               & 70134            & 76771                                & 34477                                & 37364                                \\
		$Q_\mathrm{ext}=1/\mathrm{Re}(1/(Q_\mathrm{ext,real}+iQ_\mathrm{ext,imag}))$  (Eq.\eqref{eq:real_fit_S21})                & 3226             & 499                                  & 480                                  & 2743                                 \\
		$\kappa_\mathrm{int}/2\pi$ (MHz)      & 0.217            & 0.4                                  & 1.29                                 & 1.96                                 \\
		$\kappa_\mathrm{ext}/2\pi$ (MHz)      & 4.71             & 61.5                                 & 92.4                                 & 2.66                                 \\ \hline
		Input power (dBm)                & -77              & -72                                  & -72                                  & -72                                  \\
		$n_\mathrm{cav}$                 & 4.83$\times10^6$ & \multicolumn{1}{r}{6.25$\times10^5$} & \multicolumn{1}{r}{2.84$\times10^5$} & \multicolumn{1}{r}{5.33$\times10^5$} \\ \hline
		$Q_\mathrm{int}$ (3-dB bandwidth from the Lorentzian fit) &      59789            &      75658                                &    34891                                  &   38642                                   \\ \hline
	\end{tabular}
\end{table}

\section{Supplementary Notes}

\subsection*{Microwave power dependence of the quality factor}
\begin{figure}[tb]
	\centering
	\includegraphics[width=9cm,clip]{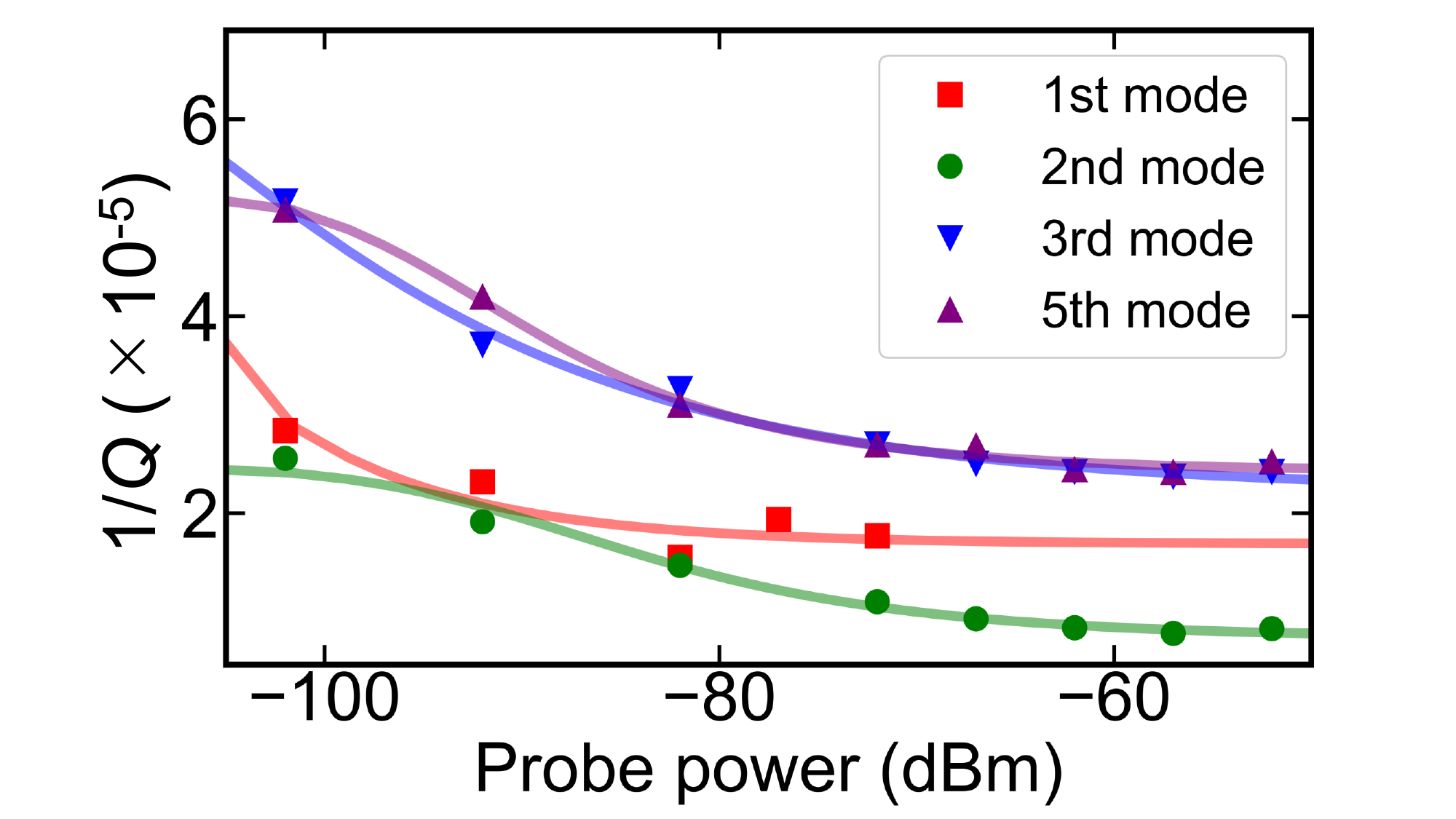}
	\caption{Microwave power dependence of the inverse of the internal quality factor measured for the first, second, third, and fifth modes. The lines are the fit using eq. \eqref{eq:saturation}.}
	\label{fig:mw_power_Q}
\end{figure}

At low probe microwave powers, inverse of the internal quality factor, $1/Q_\mathrm{int}$,  increases owing to two-level-system (TLS) losses (Fig. \ref{fig:mw_power_Q}) \cite{Gao2008}.  All the measurements in the main text are performed in the high-power regime, where the TLS loss saturation is observed. The fitting function for the data is eq. \eqref{eq:saturation}, which is discussed in the section below.

\subsection*{Temperature Dependence of the Resonance Frequency and Its Relation to the Fitting Functions}

\begin{figure}[tb]
	\centering
	\includegraphics[width=10cm,clip]{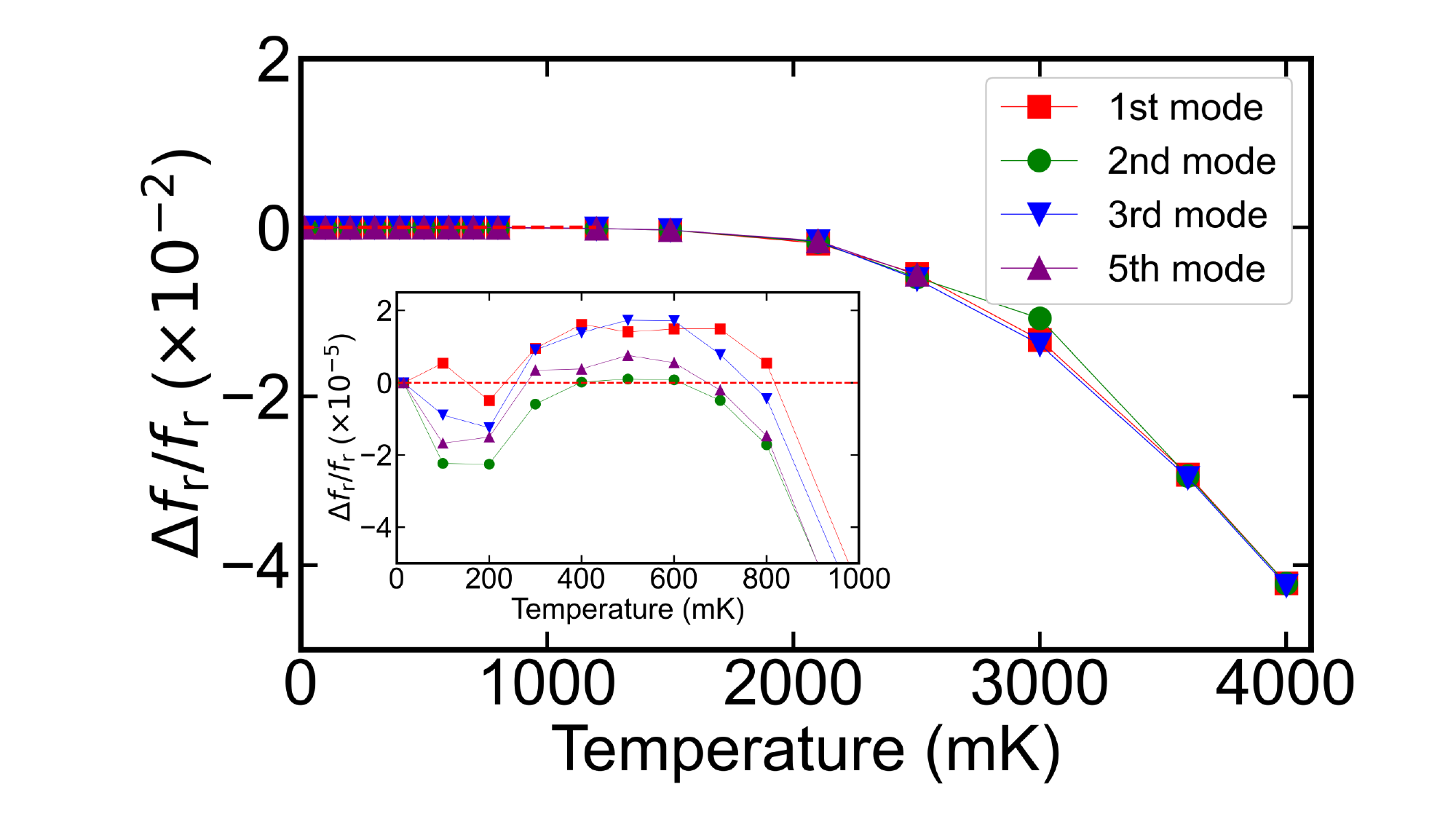}
	\caption{Temperature dependence of the relative frequency shift for each resonance mode. The inset shows the shifts below 1000 mK. The lines between dots are guides for the eye.}
	\label{fig:temp-freq}
\end{figure} 

Figure~\ref{fig:temp-freq} shows the temperature dependence of the relative frequency shift, \(\Delta f_\mathrm{r}/f_\mathrm{r}\) of the high-$L_\mathrm{k}$ resonator. At higher temperatures above 1 K, \(\Delta f_\mathrm{r}/f_\mathrm{r}\) decreases exponentially, consistent with the empirical temperature dependence of the magnetic penetration depth \cite{TinkhamBook},
\begin{equation}\label{eq:penetration_depth}
\lambda(T)=\frac{\lambda(0)}{\sqrt{1-(T/T_\mathrm{c})^4}},
\end{equation}
where \(T_\mathrm{c}\) is the superconducting critical temperature. Assuming that the change \(\Delta\lambda(T)=\lambda(T)-\lambda(0)\) is small compared to \(\lambda(0)\) ($\lambda(T)\simeq\lambda(0)$), the relative frequency shift can be related to \(\Delta\lambda(T)\) as
\begin{equation}\label{eq:df_dlambda}
	\begin{aligned}
		\frac{f_\mathrm{r}(T)-f_\mathrm{r}(T_0)}{f_\mathrm{r}(T_0)} 
		&\simeq -\frac{L_\mathrm{k,l}(T)-L_\mathrm{k,l}(T_0)}{2L_\mathrm{t,l}(T_0)} \\
		&=-\frac{1}{2L_\mathrm{t,l}(T_0)}\frac{\mu_0}{dw}\Bigl(\lambda^2(T)-\lambda^2(T_0)\Bigr) \\
		&\simeq -\frac{1}{L_\mathrm{t,l}(T_0)}\frac{\mu_0}{dw}\,\lambda(T_0)\Delta\lambda(T),
	\end{aligned}	
\end{equation}
where $L_\mathrm{t,l}$ is the total inductance per unit length, $L_\mathrm{k,l}$ is the kinetic inductance per unit length, $\mu_0$ is the vacuum permeability, $d$ is the film thickness, and $w$ is the nanowire width. 
The first equation can be derived assuming the fundamental resonance frequency of the half-wavelength transmission line resonator, $f_\mathrm{r}(T)=1/(2l\sqrt{L_\mathrm{t,l}(T)C_\mathrm{t,l}})$ \cite{Goppl2008}, where $C_\mathrm{t,l}$ is the total capacitance per unit length of the resonator and $l$ is the length of the resonator. The first equation can be applied to other resonance modes like the  quarter-wavelength resonance and higher-order resonance modes. In the second equation, we assume a one-dimensional superconducting nanowire in which the current density is uniform. For such a nanowire with width \(w\) and thickness \(d\), the kinetic inductance is given by \cite{Meservey1969}
\[
L_\mathrm{k,l} = \frac{\mu_0\lambda^2}{dw}.
\]
For a superconductor with nonuniform current distribution, the geometric factor is multiplied to the equation above \cite{Clem2013}. From eqs. \eqref{eq:penetration_depth} and  \eqref{eq:df_dlambda}, it is clearly seen that the resonance frequency decreases with increasing temperature. Additionally, when $L_\mathrm{k,l}/L_\mathrm{t,l}\simeq1$ and $n_\mathrm{s}>>n_\mathrm{qp}$, Eq. \eqref{eq:df_dlambda} can be rewritten as 
\begin{equation}
\begin{aligned}
\frac{\Delta f_\mathrm{r}(T)}{f_\mathrm{r}(T_0)}
&\simeq-\frac{L_\mathrm{k,l}(T)-L_\mathrm{k,l}(T_0)}{2L_\mathrm{k,l}(T_0)}
\\
&=\frac{n_\mathrm{s}(T)-n_\mathrm{s}(T_0)}{2n_\mathrm{s}(T)}
\\
&=
-\frac{\Delta n_\mathrm{qp}(T)}{2n_\mathrm{s}(T)},
\end{aligned}
\end{equation}
where changes in quasiparticle density is defined to be $-\Delta n_\mathrm{qp}(T) = n_\mathrm{s}(T)-n_\mathrm{s}(T_0)$ assuming $n_\mathrm{qp} + n_\mathrm{s}=\mathrm{const.}$ We also used the relation, $\lambda^2=m/(\mu_0n_\mathrm{s}e^2)$ \cite{TinkhamBook}, for the above derivation.

In contrast, at temperatures below 1 K, a small kink is observed for each mode, which is typically attributed to changes in permittivity induced by TLSs surrounding the resonator \cite{Gao2008,Muller2019,Wang2020}. According to the cavity perturbation theory \cite{PozarBook}, a small perturbation in the electromagnetic field around the resonator due to changes in permittivity shifts the resonance frequency as
\begin{equation}\label{eq:cavity_perturbation}
	\begin{aligned}
		\frac{f_\mathrm{r}'-f_\mathrm{r}}{f_\mathrm{r}}
		&\simeq -\frac{\int_{W_\mathrm{local}} d\bm{r}\,\mathrm{Re}(\Delta\tilde{\epsilon})\,|\bm{E}_0|^2}{\int_{W_0} d\bm{r}\,\Bigl(\mathrm{Re}(\tilde{\epsilon})|\bm{E}_0|^2+\mathrm{Re}(\tilde{\mu})|\bm{H}_0|^2\Bigr)} \\
		&\simeq -\mathrm{Re}(\Delta\tilde{\epsilon})\,\frac{\int_{W_\mathrm{local}} d\bm{r}\,|\bm{E}_0|^2}{\int_{W_0} d\bm{r}\,\Bigl(\mathrm{Re}(\tilde{\epsilon})|\bm{E}_0|^2+\mathrm{Re}(\tilde{\mu})|\bm{H}_0|^2\Bigr)} \\
		&=-p(W_\mathrm{local})\mathrm{Re}(\Delta\tilde{\epsilon}),
	\end{aligned}
\end{equation}
where $f_\mathrm{r}$ is the unperturbed resonance frequency, $f_\mathrm{r}'$ is the perturbed resonance frequency, \(W_\mathrm{local}\) is the region in which the permittivity changes, \(W_0\) is the entire region containing the resonator’s electromagnetic field, \(\bm{E}_0\) and \(\bm{H}_0\) are the unperturbed electric and magnetic fields, respectively, \(\Delta \tilde{\epsilon}\) represents the change in complex permittivity, \(\tilde{\epsilon}\) is the complex permittivity, and \(\tilde{\mu}\) is the complex permeability. Here,
\[
p(W_\mathrm{local})=\frac{\int_{W_\mathrm{local}} d\bm{r}\,|\bm{E}_0|^2}{\int_{W_0} d\bm{r}\,\Bigl(\mathrm{Re}(\tilde{\epsilon})|\bm{E}_0|^2+\mathrm{Re}(\tilde{\mu})|\bm{H}_0|^2\Bigr)}
\]
is the participation ratio or the filling factor corresponding to the fraction of the volume in which the permittivity change occur. We assume that the spatial variation in \(\Delta\tilde{\epsilon}\) is negligible within \(W_\mathrm{local}\). This assumption is reasonable for both the global, temperature-driven changes in \(\epsilon\) and the local, laser-induced changes in \(\epsilon\), given the homogeneous substrate and TLS distribution.

Based on the standard tunneling model (STM) for TLS \cite{Phillips1987,Gao2008}, the temperature dependence of the TLS permittivity is written as
\begin{equation}\label{eq:epsilon_change_TLS}
	\begin{aligned}
		\mathrm{Re}(\Delta\tilde{\epsilon}_\mathrm{TLS})&= \mathrm{Re}(\tilde{\epsilon}_\mathrm{TLS}(T))-\mathrm{Re}(\tilde{\epsilon}_\mathrm{TLS}(0)) \\
		&=-\frac{\delta_\mathrm{TLS}}{\pi}\left[\Re\Psi\left(\frac{1}{2}-\frac{hf_\mathrm{r}}{2i\pi k_BT}\right)-\log\left(\frac{hf_\mathrm{r}}{2\pi k_BT}\right)\right],
	\end{aligned}
\end{equation}
where \(\delta_\mathrm{TLS}\) is the intrinsic TLS loss, \(\Psi(z)\) is the digamma function, \(\omega_\mathrm{r}\) is the angular resonance frequency of the resonator, \(k_B\) is the Boltzmann constant, and \(h\) is the Planck constant.

\begin{figure}[b]
	\centering
	\includegraphics[width=14cm,clip]{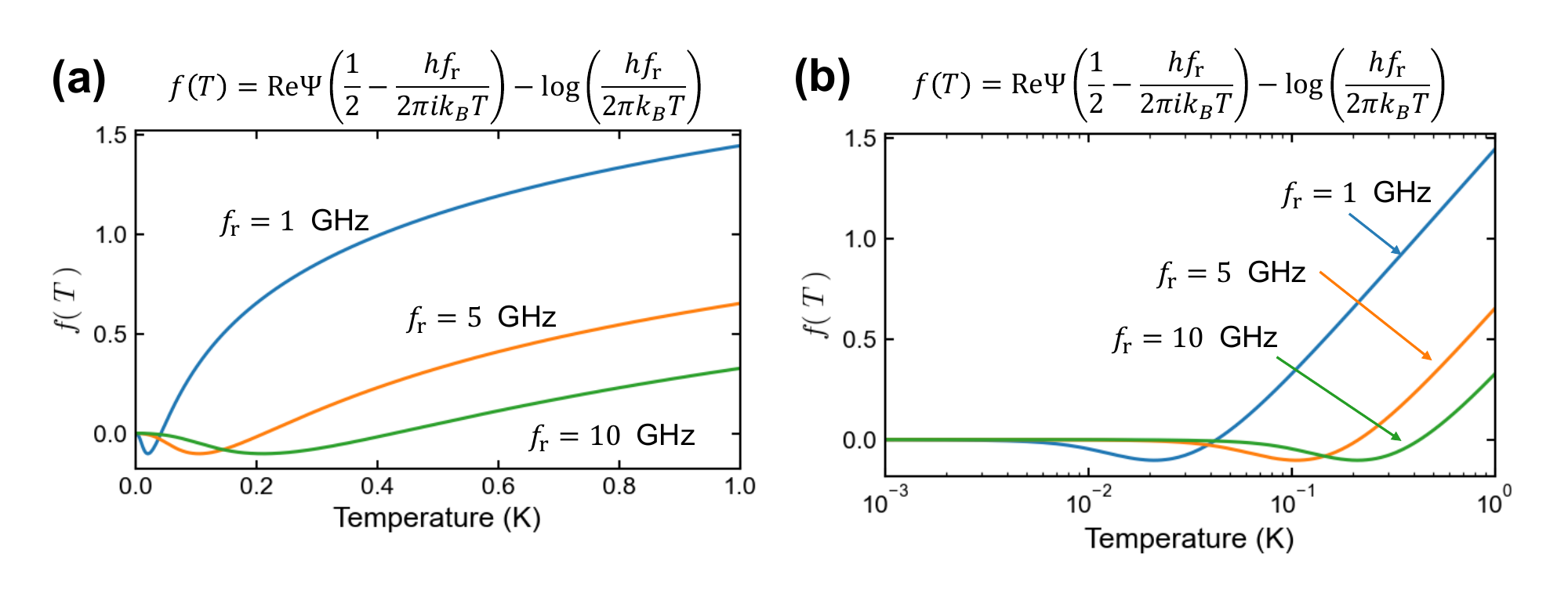}
	\caption{(a) Temperature-dependent terms in Eq.~\eqref{eq:freq_shift_final} with different resonator frequencies, \(f_\mathrm{r}\); (b) the corresponding logarithmic plot. (c) Temperature dependence of the digamma function; (d) the logarithmic term; (e) and (f) Functions used to fit the experimental data in the main text.}
	\label{fig:digamma_log_sim}
\end{figure}

From Eqs.~\eqref{eq:cavity_perturbation} and \eqref{eq:epsilon_change_TLS}, the relative frequency shift can be expressed as
\begin{equation}\label{eq:freq_shift_final}
	\frac{\Delta f_\mathrm{r}(T)}{f_\mathrm{r}} = p(W_\mathrm{local})\frac{\delta_\mathrm{TLS}}{\pi}\left[\Re\Psi\left(\frac{1}{2}-\frac{hf_\mathrm{r}}{2i\pi k_BT}\right)-\log\left(\frac{hf_\mathrm{r}}{2\pi k_BT}\right)\right].
\end{equation}
Figure~\ref{fig:digamma_log_sim}(a)(b) shows the functional forms of Eq.~\eqref{eq:freq_shift_final} for different resonator frequencies. The downward shift at lower temperatures (\(k_BT < hf_\mathrm{r}\)) can be attributed to the digamma function, while the upward shift originates from the logarithmic term. 

\subsection*{Temperature and Optical Power Dependence of the Resonance Frequency}

\begin{figure}[tb]
	\centering
	\includegraphics[width=13.5cm,clip]{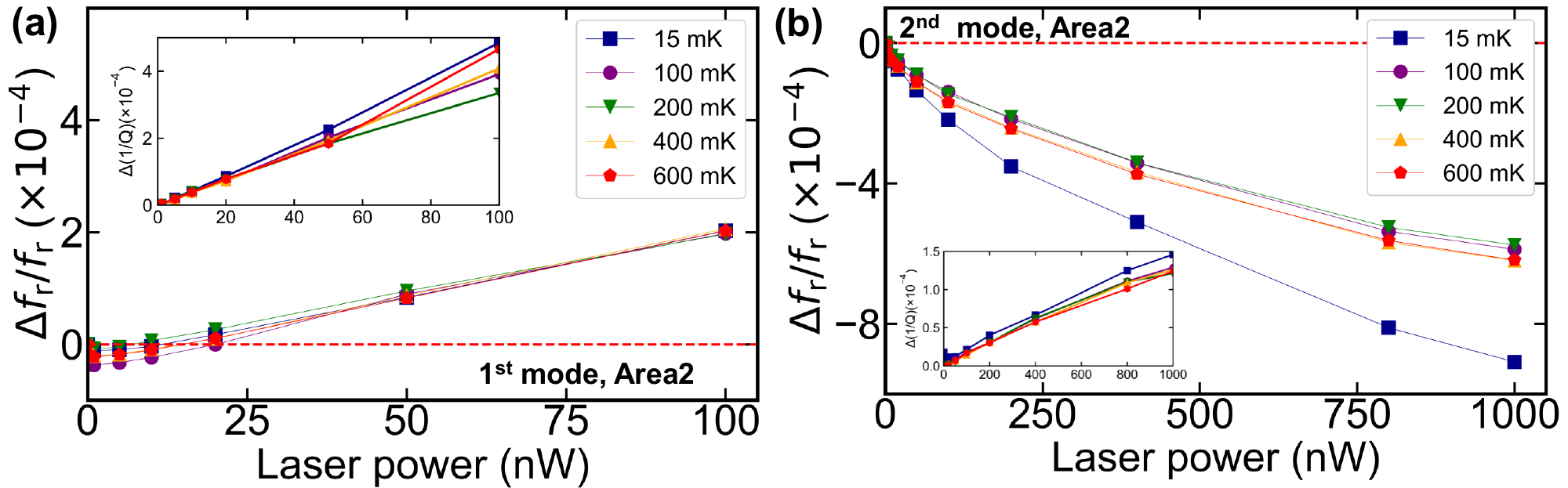}
	\caption{Optical power dependence of the relative frequency shifts at different temperatures, taken at (a) the first resonance mode and Area 2, and (b) the second resonance mode and Area 2. The lines between dots are guide for eye. The inset shows $\Delta(1/Q)\ (P_\mathrm{opt})$ at these temperatures.}
	\label{fig:temp-opt}
\end{figure} 

To distinguish between equilibrium temperature increases and the effects of laser irradiation, we measure the laser-induced frequency shifts at different temperatures. Figure~\ref{fig:temp-opt} shows the laser-power dependence of \(\Delta f_\mathrm{r}/f_\mathrm{r}\) at mixing-chamber temperatures from 15--600~mK. The overall trend of the frequency shifts, regardless of sign, remains unchanged with increasing temperature. Additionally, the temperature-induced relative frequency shifts are on the order of \(10^{-5}\) for all resonant modes (15--600~mK in Fig.~\ref{fig:temp-freq}), whereas the laser-induced shifts are on the order of \(10^{-4}\) (Fig.~\ref{fig:temp-opt}). In other words, even at higher temperatures \(\bigl(T \sim h f_\mathrm{r}/k_\mathrm{B}\bigr)\), the laser-induced effects have a larger impact on the resonance frequencies than the temperature-induced changes in TLS states. These differences indicate that the laser-induced changes in the environment are distinct from those resulting from an equilibrium temperature increase. We attribute the larger changes observed under laser irradiation to high-energy TLSs that are coupled to the resonator and excited by nonequilibrium phonons.

Furthermore, as shown in the inset of Fig.~\ref{fig:temp-opt}, the values of \(\Delta(1/Q)\) are also insensitive to equilibrium temperature for the first and second modes at Area~2. Nearly resonant TLSs with \(f_\mathrm{TLS}\sim 2\text{--}5~\mathrm{GHz}\) are excited at these temperatures. If resonant TLSs were the primary contributors to the loss increase, \(\Delta(1/Q)\) would be temperature dependent. The temperature-independent \(\Delta(1/Q)\) suggests that resonant TLSs do not play a dominant role in the increase in loss. Instead, high-energy TLSs \(\bigl(f_\mathrm{TLS}>f_\mathrm{r}\bigr)\) can increase the resonator loss.

\subsection*{Microwave and Optical Power Dependence of the Resonance Frequency}

\begin{figure}[tb]
	\centering
	\includegraphics[width=14cm,clip]{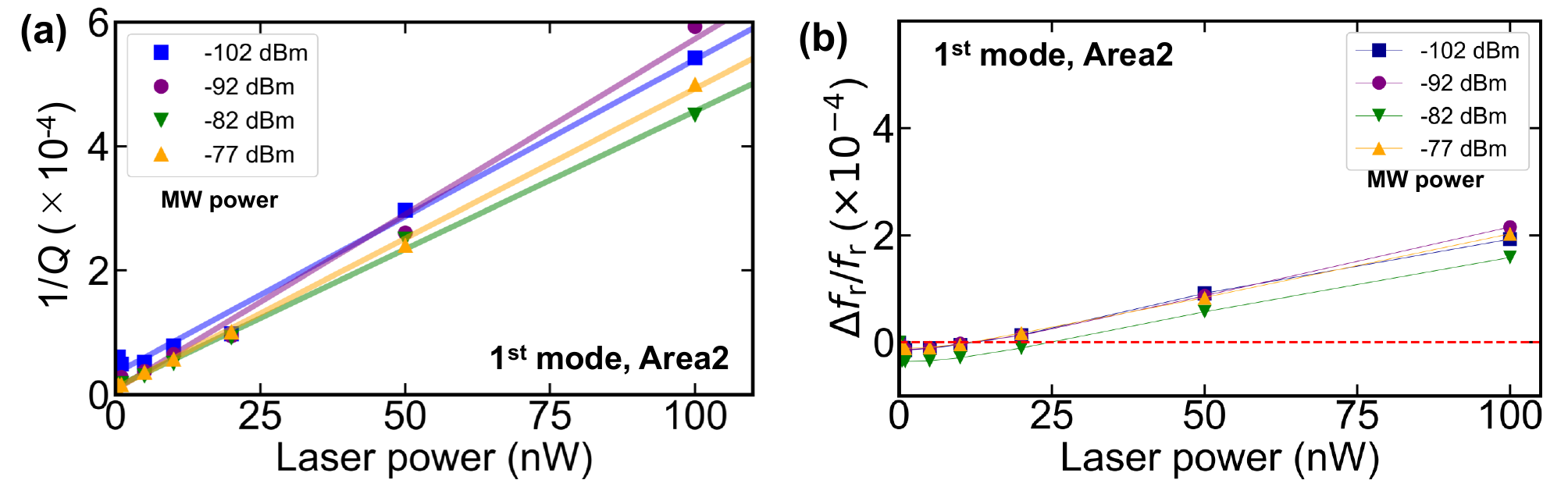}
	\caption{Optical power dependence of (a) \(1/Q\) and (b) the relative frequency shift, measured at different probe microwave powers (MW power). The lines are the linear curve fit for $1/Q$, while those in $\Delta f_\mathrm{f}/f_\mathrm{r}$ between dots are guide for eye.}
	\label{fig:mw_power_freq}
\end{figure} 

We also investigate the microwave-power dependence of the relative frequency shift to assess the effects of TLS saturation under high-power microwaves. Figure~\ref{fig:mw_power_freq} shows the laser-induced frequency shifts measured at different probe microwave powers. The highest power used, $-$77 dBm, is approximately the TLS saturation power. The upward frequency shift remains unchanged as the microwave power decreases. Thus, in this power range, microwave-induced saturation does not impact the laser-induced shifts for the first mode. It is consistent with our theoretical model considering the contribution from high-energy  non-resonant TLSs.


\subsection*{Derivation of the Permittivity of TLSs}\label{subsec:TLS_permittivity}

In this section, we briefly summarize the relationship between the real and imaginary parts of the complex permittivity arising from TLSs \cite{Phillips1987,Gao2008}.
A different treatment for the complex permittivity of TLSs is introduced in Refs. \cite{Faoro2012,Burnett2016}.
According to the standard tunneling model (STM), the dielectric loss due to TLSs can be written as
\begin{equation}\label{eq:TLS-loss-temp}
	\mathrm{Im}(\tilde{\epsilon}_\mathrm{TLS}(T,f)) = \delta_\mathrm{TLS}\,\tanh\qty(\frac{hf}{2k_\mathrm{B}T}).
\end{equation}
Under high driving powers, considering the saturation of TLSs, this expression is modified to \cite{Phillips1987,Gao2008}
\begin{equation}\label{eq:saturation}
	\mathrm{Im}(\tilde{\epsilon}_\mathrm{TLS}(T,f,P_\mathrm{mw})) = \delta_\mathrm{TLS}\,\tanh\qty(\frac{hf}{2k_\mathrm{B}T})\frac{1}{\sqrt{1+\left(\frac{n_\mathrm{cav} }{n_\mathrm{c}}\right)^\beta}},
\end{equation}
where \(P_\mathrm{mw}\) is the input microwave power, \( n_\mathrm{cav} \) (which is proportional to \(P_\mathrm{mw}\)) is the average number of photons in the resonator, \(n_\mathrm{c}\) is the critical photon number determined by the TLS relaxation and decoherence times and their coupling to the resonator, and \(\beta\) is an empirical fitting parameter used to reproduce experimental observations \cite{Muller2019}. In addition, \(\delta_\mathrm{TLS}\) is typically expressed as \cite{Phillips1987,Gao2008,Burnett2016}
\begin{equation}\label{eq:TLS_loss_tan}
	\delta_\mathrm{TLS} = \frac{\pi \rho_\mathrm{TLS} d_0^2}{3\epsilon_\mathrm{host}},
\end{equation}
where \(\rho_\mathrm{TLS}\) is the TLS density of states, \(d_0\) is the TLS dipole moment, and \(\epsilon_\mathrm{host}\) is the permittivity of the host material. 

The real part of the complex permittivity is obtained from the Kramers--Kronig relation as follows:
\begin{equation}
	\mathrm{Re}(\tilde{\epsilon}_\mathrm{TLS}(T,f)) =
	1 + \frac{2}{\pi}\,\mathcal{P}\int_{0}^{\infty} df'\,
	\frac{f'\,\mathrm{Im}(\tilde{\epsilon}_\mathrm{TLS}(T,f))}{f'^{2} - f^{2}},
\end{equation}
where \(\mathcal{P}\) denotes that the integral is taken as the Cauchy principal value. By evaluating the integral, Eq.~\eqref{eq:epsilon_change_TLS} is obtained. Note that, since the integral is weighted by the factor \(f'^{2} - f^{2}\), variations in \(\mathrm{Im}(\tilde{\epsilon}_\mathrm{TLS}(f'))\) near the resonance frequency predominantly determine \(\mathrm{Re}(\tilde{\epsilon}_\mathrm{TLS})(T,f)\).

\section{Supplementary Discussion}

\subsection{Analysis of a coupled TLS-resonator system}
\subsubsection{Transverse interaction}
In this section, we derive the complex frequency shift of a resonator coupled to a single TLS following Refs. \cite{Kirsh2017a,Capelle2020}. First, we consider only the transverse coupling for simplicity.

The Hamiltonian of the TLS-resonator system is 
\begin{equation}
\hat{H}_\mathrm{TLS-MWR}/\hbar 
= \omega_\mathrm{TLS}\hat{\sigma}^+\hat{\sigma}
+\omega_\mathrm{r}\hat{c}^\dagger\hat{c}
+g_{\mathrm{\perp}}(\hat{\sigma}^+\hat{c}+\hat{\sigma}\hat{c}^\dagger)
,
\end{equation}
where $\hat{\sigma}\ (\hat{c})$ is the TLS lowering (resonator annihilation) operator, and $g_\perp$ is the transverse vacuum coupling between a TLS and the resonator. In the frame rotating at the resonator frequency,
\begin{equation}
\hat{H}_\mathrm{TLS-MWR}^\mathrm{rot}/\hbar 
= -\Delta_\mathrm{TLS}\hat{\sigma}^+\hat{\sigma}
+g_{\mathrm{\perp}}(\hat{\sigma}^+\hat{c}+\hat{\sigma}\hat{c}^\dagger)
,
\end{equation}
where $\Delta_\mathrm{TLS}=\omega_\mathrm{r}-\omega_\mathrm{TLS}$. 
Decoherence is incorporated using Lindblad superoperators in the framework of the quantum master equation as
\begin{equation}
\dv{\rho}{t}
= -\frac{i}{\hbar}[\hat{H},\rho] 
+ \Gamma_{\downarrow}\mathcal{D}_{\hat{\sigma}}(\rho)
+
\Gamma_{\uparrow}\mathcal{D}_{\hat{\sigma}^+}(\rho)
+
2\gamma_\phi\mathcal{D}_{\hat{\sigma}^+\hat{\sigma}}(\rho)
+
\kappa_\mathrm{tot}
\mathcal{D}_{\hat{c}}(\rho),
\end{equation}
\begin{equation}
\mathcal{D}_{\hat{L}}(\rho)
=\hat{L}\rho\hat{L}^\dagger
-\frac{1}{2}\qty{\hat{L}^\dagger\hat{L},\rho},
\end{equation}
where $[\hat{A},\hat{B}]=\hat{A}\hat{B}-\hat{B}\hat{A}$,  $\qty\big{\hat{A},\hat{B}}=\hat{A}\hat{B}+\hat{B}\hat{A}$. The time evolution of an operator is obtained using
\begin{equation}
\ev*{\hat{A}}=\mathrm{Tr}(\rho \hat{A}),\ 
\dv{\ev*{\hat{A}}}{t}=\mathrm{Tr}\qty(\hat{A}\dv{\rho}{t}),	
\end{equation}
where $\ev{\cdot}$ is the expectation value of the operator. Then, 
\begin{equation}\label{eq:dsigma_dt}
	\dv{\ev{\hat{\sigma}}}{t}
	= -\qty(-i\Delta_{\mathrm{TLS}}
	+\Gamma_2)\,\langle \hat{\sigma}\rangle
	\;+\; i g_{\mathrm{\perp}}\,\langle \hat{\sigma}_z \hat{c}\rangle,
\end{equation}	
\begin{equation}\label{eq:dc_dt}
	\dv{\ev{\hat{c}}}{t}
	= -\frac{\kappa_{\mathrm{tot}}}{2}\langle \hat{c}\rangle
	\;-\; i g_{\mathrm{\perp}}\,\langle \hat{\sigma}\rangle
	,
\end{equation}	
\begin{equation}\label{eq:dsigz_dt}
	\dv{\ev{\hat{\sigma}_z}}{t}
	= 2 i g_{\mathrm{\perp}}\,\langle \hat{\sigma}\hat{c}^\dagger-\hat{\sigma}^\dagger\hat{c}\rangle
	-\Gamma_1(\ev{\hat{\sigma}_z}-\ev{\hat{\sigma}_z}_\mathrm{0}),
\end{equation}
where $\Gamma_1=\Gamma_{\downarrow}+\Gamma_{\uparrow}=1/T_1$ ($\Gamma_{\downarrow}=\Gamma_0(n_\mathrm{p}+1)$: phonon emission rate, $\Gamma_{\uparrow}=\Gamma_0n_\mathrm{p}$: phonon absorption rate,  
$T_1$: relaxation time of TLS), $\Gamma_2=1/T_2 = 1/2T_1 + \gamma_\phi$ ($T_2$: coherence time of TLS, $\gamma_\phi$: pure dephasing of TLS), $(\Gamma_{\downarrow}-\Gamma_{\uparrow})/(\Gamma_{\uparrow}+\Gamma_{\downarrow})=1/(2n_\mathrm{p}+1)=\mathrm{tanh}(\hbar\omega_\mathrm{TLS}/2k_BT)=-\ev{\hat{\sigma}_z}_\mathrm{0}$ (population imbalance), $n_\mathrm{p}$ is the number of phonon interacting with a TLS, and $\kappa_\mathrm{tot}$ is the total loss of the resonator. Under equilibrium conditions, $n_\mathrm{p}$ is the Bose-Einstein occupation.

Assuming the absence of correlation between $\ev{\hat{c}}$, $\ev{\hat{\sigma}}$, and $\ev{\hat{\sigma}_z}$ (i.e. $\ev{\hat{\sigma}^+\hat{c}}=\ev{\hat{\sigma}^+}\ev{\hat{c}}$), and the steady state such that the time derivative terms are zero, Eq.\eqref{eq:dsigma_dt} gives
\begin{equation}
	\ev{\hat{\sigma}}
	=\frac{ig_{\mathrm{\perp}}\ev{\hat{\sigma}_z}\ev{\hat{c}}}{\Gamma_2-i\Delta_{\mathrm{TLS}}}.
\end{equation}
Substituting above equation to Eq.\eqref{eq:dc_dt} yields 
\begin{equation}
	\dv{\ev{\hat{c}}}{t}
	=\qty(-\frac{\kappa_\mathrm{tot}}{2}+\frac{g_{\mathrm{\perp}}^2\ev{\hat{\sigma}_z}}{\Gamma_2-i\Delta_{\mathrm{TLS}}})\ev{\hat{c}}.
\end{equation}
Thus, additional loss due to a single TLS is 
\begin{equation}\label{eq:single-TLS-loss}
	\delta\kappa
	=-\frac{2g_{\mathrm{\perp}}^2\Gamma_2\ev{\hat{\sigma}_z}}{\Gamma_2^2+\Delta_{\mathrm{TLS}}^2},
\end{equation}
and frequency shift due to a single TLS is 
\begin{equation}\label{eq:single-TLS-df}
	\delta\omega
	=-\frac{g_{\mathrm{\perp}}^2\Delta_{\mathrm{TLS}}\ev{\hat{\sigma}_z}}{\Gamma_2^2+\Delta_{\mathrm{TLS}}^2}.
\end{equation}
Furthermore, from Eq. \eqref{eq:dsigz_dt},
\begin{equation}
	\begin{aligned}
		\ev{\hat{\sigma}_z}
		=\frac{\ev{\hat{\sigma}_z}_\mathrm{0}}{1+
			\frac{n_\mathrm{cav}}{n_s}
			\frac{\Gamma_2^2}{\Gamma_2^2+\Delta_{\mathrm{TLS}}^2}
		}
		,
	\end{aligned}
\end{equation}
where $n_\mathrm{cav}=|\ev{\hat{c}}|^2$ and  $n_s=\Gamma_1\Gamma_2/4g_{\mathrm{\perp}}^2$. It is evident that $\ev{\hat{\sigma}_z}\leqq\ev{\hat{\sigma}_z}_0$.

\subsubsection{Longitudinal coupling}
Then, we consider the loss and frequency shift due to the longitudinal coupling between a TLS and the resonator.
\begin{equation}
	\hat{H}_\mathrm{TLS-MWR}/\hbar 
	= \omega_\mathrm{TLS}\hat{\sigma}^+\hat{\sigma}
	+\omega_\mathrm{r}\hat{c}^\dagger\hat{c}
	+g_{\mathrm{\parallel}}\hat{\sigma}_z(\hat{c}^\dagger+\hat{c})
	,
\end{equation}
where $g_{\mathrm{\parallel}}$ is the longitudinal coupling constant. Similar to the treatment above, the dynamics of the operators can be calculated as
\begin{equation}\label{eq:dc_dt_para}
	\dv{\ev{\hat{c}}}{t}
	=-(i\omega_\mathrm{r}+\kappa_{\mathrm{tot}}/2)\ev{\hat{c}}-ig_{\parallel}\ev{\hat{\sigma}_z},
\end{equation}
\begin{equation}
	\dv{\ev{\hat{\sigma}}}{t}
	=-(i\omega_\mathrm{TLS}+\Gamma_2)\ev{\hat{\sigma}}-2ig_{\parallel}\ev{\hat{\sigma}(\hat{c}^\dagger+\hat{c})},
\end{equation}
\begin{equation}\label{eq:dsigz_para}
	\dv{\ev{\hat{\sigma}_z}}{t}
	=-\Gamma_1(\ev{\hat{\sigma}_z}-\ev{\hat{\sigma}_z}_{0}).
\end{equation}
Since
\begin{equation}
	\dv{\ev{\hat{\sigma}}}{t}
	=-\qty{i[\omega_\mathrm{TLS}+2g_{\parallel}(\ev{\hat{c}}^*+\ev{\hat{c}})]+\Gamma_2}\ev{\hat{\sigma}},	
\end{equation}
frequency shift of TLS due to the longitudinal coupling is $2g_{\parallel}(\ev{\hat{c}}^*+\ev{\hat{c}})$ ($\omega_\mathrm{TLS}\rightarrow\omega_\mathrm{TLS}+2g_{\parallel}(\ev{\hat{c}}^*+\ev{\hat{c}})$). Substituting this into  $\ev{\hat{\sigma}_z}_0=-\mathrm{tanh}(\hbar\omega_\mathrm{TLS}/2k_BT)$ yields
\begin{equation}
	\begin{aligned}
		\ev{\hat{\sigma}_z}_0
		&=-\mathrm{tanh}\qty(\frac{\hbar\omega_\mathrm{TLS}+\hbar[2g_{\parallel}(\ev{\hat{c}}^*+\ev{\hat{c}})]}{2k_BT})
		\\
		&=-\mathrm{tanh}\qty(\frac{\hbar\omega_\mathrm{TLS}}{2k_BT})-\qty[1-\mathrm{tanh}^2\qty(\frac{\hbar\omega_\mathrm{TLS}}{2k_BT})]\frac{\hbar g_{\parallel}(\ev{\hat{c}}^*+\ev{\hat{c}})}{k_BT},
	\end{aligned}
\end{equation}
indicating that the population imbalance is modulated by the TLS-resonator coupling. We use $\hbar\omega_\mathrm{TLS}>>\hbar g_\parallel$ in the second line. Assuming $\ev{\hat{c}(t)}=\ev{\hat{c}(t)}_0e^{-i\omega_\mathrm{r} t}$, $\ev{\hat{\sigma}_z}=s_z+\delta s_ze^{-i\omega_\mathrm{r} t}$, $\ev{\hat{\sigma}_z}_0=s_{z0}+\delta s_{z0}\ev{\hat{c}(t)}_0 e^{-i\omega_\mathrm{r} t}+\delta s_{z0}\ev{\hat{c}(t)}^*_0 e^{i\omega_\mathrm{r} t}$ in Eq. \eqref{eq:dsigz_para} ($s_{z0}=-\mathrm{tanh}(\hbar\omega_\mathrm{TLS}/2k_BT)$, $\delta s_{z0}=-\hbar g_\parallel[1-\mathrm{tanh}^2(\hbar\omega_\mathrm{TLS}/2k_BT)]/k_BT$) and equating terms rotating with $e^{-i\omega_\mathrm{r}t}$, 
\begin{equation}
	s_z = s_{z0},\ 
	\delta s_z = 
	\frac{\Gamma_1\delta s_{z0}\ev{\hat{c}(t)}_0}{\Gamma_1-i\omega_\mathrm{r}}. 
\end{equation}
Substituting $\ev{\hat{\sigma}_z}$ to Eq. \eqref{eq:dc_dt_para} and focusing on terms rotating with $e^{-i\omega_\mathrm{r}t}$ lead to
\begin{equation}
	\begin{aligned}
		\dv{\ev{\hat{c}(t)}_0}{t}
		&=-\qty(\frac{\kappa_{\mathrm{tot}}}{2})\ev{\hat{c}(t)}_0
		-ig_{\parallel}\frac{\Gamma_1}{\Gamma_1-i\omega_\mathrm{r}}\delta s_{z0}\ev{\hat{c}(t)}_0
		\\
		&=
		-\qty{
			\frac{\kappa_{\mathrm{tot}}}{2}
			-i\qty[1-\mathrm{tanh}^2\qty(\frac{\hbar\omega_\mathrm{TLS}}{2k_BT})]
			\frac{\Gamma_1}{\Gamma_1-i\omega_\mathrm{r}}
			\frac{\hbar g_{\parallel}^2}{k_BT}
		}\ev{\hat{c}(t)}_0.
	\end{aligned}
\end{equation}
Thus, the complex frequency shift due to the longitudinal coupling is 
\begin{equation}
\qty[1-\mathrm{tanh}^2\qty(\frac{\hbar\omega_\mathrm{TLS}}{2k_BT})]
\frac{\Gamma_1\omega_\mathrm{r}}{\Gamma_1^2+\omega_\mathrm{r}^2}
\frac{2\hbar g_{\parallel}^2}{k_BT},
\end{equation}
for the TLS-induced loss and
\begin{equation}
	-
	\qty[1-\mathrm{tanh}^2\qty(\frac{\hbar\omega_\mathrm{TLS}}{2k_BT})]
	\frac{\Gamma_1^2}{\Gamma_1^2+\omega_\mathrm{r}^2}
	\frac{\hbar g_{\parallel}^2}{k_BT},
\end{equation}
for the TLS-induced frequency shift.

For the general, nonequilibrium case such that $\ev{\hat{\sigma}_z}_0\neq-\mathrm{tanh}(\hbar\omega_\mathrm{TLS}/2k_BT)$ and  $\ev{\hat{\sigma}_z}_0=\ev{\hat{\sigma}_z(\omega)}_0$, 
\begin{equation}
\ev{\hat{\sigma}_z(\omega_\mathrm{TLS}+\delta\omega)}_0
\simeq
\ev{\hat{\sigma}_z(\omega_\mathrm{TLS})}_0
+\left.\dv{\ev{\hat{\sigma}_z}_0}{\omega}\right|_{\omega=\omega_\mathrm{TLS}}
\hbar g_\parallel(\ev{\hat{c}}^*+\ev{\hat{c}}),
\end{equation}
where $\delta\omega=g_\parallel(\ev{\hat{c}}^*+\ev{\hat{c}})$. Then, $[1-\mathrm{tanh}^2(\hbar\omega_\mathrm{TLS}/2k_BT)]/k_BT\rightarrow(\mathrm{d}\ev{\hat{\sigma}_z}_0/\mathrm{d}\omega)|_{\omega=\omega_\mathrm{TLS}}$. It should be noted that we assumes the rotating wave approximation in the above treatment. Additionally, similar expressions can be found in Ref. \cite{Gao2008}.

\subsubsection{Total loss from all TLSs}
For simplicity, we assume $\ev{\hat{\sigma}_{z}}=\ev{\hat{\sigma}_{z}}_0=-\mathrm{tanh}(\hbar\omega_\mathrm{TLS}/2k_BT)$ by neglecting the saturation effect due to a microwave probe tone in the following analysis. Furthermore, we use symbols $\ev{\hat{\sigma}_{z}}_0 \rightarrow S\ (-1\leqq S\leqq0)$ and
$[1-\mathrm{tanh}^2(\hbar\omega_\mathrm{TLS}/2k_BT)]\hbar/k_BT \rightarrow\mathrm{d}S\ (0\leqq\mathrm{d}S)$ 
as parameters representing the degree of nonequilibrium excitation (population imbalance) independent of equilibrium temperature, $T$. Integrating Eq.~\eqref{eq:single-TLS-loss} over all TLSs within the effective volume yields
\begin{equation}\label{eq:delta_kappa_tot}
\begin{aligned}
\delta\kappa_{\mathrm{TLS,total}}
&=\sum_{i}\qty(-
\frac{2g_{i,\mathrm{\perp}}^2\Gamma_{i,2}S_i}{\Gamma_{i,2}^2+\Delta_{i,\mathrm{TLS}}^2}
+
\mathrm{d}S_i
\frac{\Gamma_{i,1}\omega_\mathrm{r}}{\Gamma_{i,1}^2+\omega_\mathrm{r}^2}
2 g_{i,\parallel}^2
)
\\
&\simeq-2\tilde{g}_{\mathrm{\perp}}^2\tilde{S}
\int_{\Delta_\mathrm{min}}^{\Delta_\mathrm{max}}
\hbar\rho_\mathrm{TLS}V_\mathrm{eff}\mathrm{d}\Delta_{\mathrm{TLS}}
\frac{\tilde{\Gamma}_2}{\tilde{\Gamma}_2^2+\Delta_{\mathrm{TLS}}^2}
+\mathrm{d}\tilde{S}
\frac{\tilde{\Gamma}_{1}\omega_\mathrm{r}}{\tilde{\Gamma}_{1}^2+\omega_\mathrm{r}^2}
2\tilde{g}_{\parallel}^2
\int_0^{\omega_\mathrm{max}}\hbar\rho_\mathrm{TLS}V_\mathrm{eff}\mathrm{d}\omega
\\
&=
\underbrace{-2\pi\hbar\rho_\mathrm{TLS}V_\mathrm{eff}\tilde{g}_{\mathrm{\perp}}^2\tilde{S}}_{\text{Resonant loss due to the transverse coupling}}
\underbrace{+2\hbar\rho_{\mathrm{TLS}}V_\mathrm{eff} \tilde{g}_{\parallel}^2
	\frac{\tilde{\Gamma}_{1}\omega_\mathrm{r}}{\tilde{\Gamma}_{1}^2+\omega_\mathrm{r}^2}
	\omega_\mathrm{max}\mathrm{d}\tilde{S}}_
	{\text{Debye (non-resonant) loss due to the longitudinal coupling}}
,
\end{aligned}
\end{equation}
Here, index $i$ corresponds to the $i$th TLS,  $\rho_\mathrm{TLS}(\omega)=\rho_\mathrm{TLS}$ (J$^{-1}$m$^{-3}$) is the energy-independent TLS density of states, and $V_\mathrm{eff}$ is the relevant volume around the resonator. 
In the following analysis, we consider that TLSs' parameters are represented by a single, averaged value for all relevant TLSs using $\tilde{\cdot}$. In other words, we neglect the frequency dependence and individual differences of these parameters. Also, we assume that $|\Delta_\mathrm{max}|>>\omega_\mathrm{r}=|\Delta_\mathrm{min}|>>\Gamma_2$. At sufficiently low temperatures such that $\tilde{S}=-1$ and $\mathrm{d}\tilde{S}=0$, the TLS-induced loss is 
\begin{equation}
	\delta\kappa_{\mathrm{TLS,total}}=2\pi\hbar\rho_{\mathrm{TLS}}V_\mathrm{eff}\tilde{g}_\perp^2,
\end{equation}
which corresponds to Eq. \eqref{eq:TLS_loss_tan} in the macroscopic picture. For experiments at equilibrium conditions (low temperatures satisfying $\tilde{S}\simeq-1$ for a broad frequency range), the second term in Eq. \eqref{eq:delta_kappa_tot} is usually negligible.

\subsubsection{Total frequency shift due to all TLSs}

For the resonator frequency shift, we also perform integration for all relevant TLSs as 
\begin{equation}\label{eq:delta_f_tot}
	\begin{aligned}
		\delta\omega_\mathrm{TLS,total}
		&=\sum_{i}\qty(
		-\frac{g^2\Delta_{i,\mathrm{TLS}}S_i}{\Gamma_{i,2}^2+\Delta_{i,\mathrm{TLS}}^2}
		-
		\mathrm{d}S_i
		\frac{\Gamma_{i,1}^2}{\Gamma_{i,1}^2+\omega_\mathrm{r}^2}
		 g_{i,\parallel}^2
		)
		\\
		&\simeq
		g_\perp^2\tilde{S}
		\int_{\Delta_{\mathrm{min}}'}^{\Delta_\mathrm{max}'}\hbar\rho_\mathrm{TLS}V_\mathrm{eff}\mathrm{d}\Delta_{\mathrm{TLS}}
		\frac{\Delta_{\mathrm{TLS}}}{\tilde{\Gamma}_2^2+\Delta_{\mathrm{TLS}}^2}
		-
		\mathrm{d}\tilde{S}
		\frac{\tilde{\Gamma}_{1}^2}{\tilde{\Gamma}_{1}^2+\omega_\mathrm{r}^2}
		\tilde{g}_{\parallel}^2
		\int_{0}^{\omega_\mathrm{max}}
		\hbar\rho_{\mathrm{TLS}}V_\mathrm{eff}\mathrm{d}\omega
		\\
		&\simeq
		\underbrace{\hbar\rho_{\mathrm{TLS}}V_\mathrm{eff}
			\tilde{g}_{\perp}^2
			\mathrm{ln}\qty(\frac{\Delta_\mathrm{max}'}{\Delta_{\mathrm{min}}'})
			\tilde{S}}_
			{\text{Freq. shift due to the transverse coupling}}
		\underbrace{-\hbar\rho_{\mathrm{TLS}}V_\mathrm{eff}
			\tilde{g}_{\parallel}^2
			\frac{\tilde{\Gamma}_{1}^2}{\tilde{\Gamma}_{1}^2+\omega_\mathrm{r}^2}
			\omega_\mathrm{max}
			\mathrm{d}\tilde{S}
		}_{\text{Freq. shift due to the longitudinal coupling}}
		.
	\end{aligned}
\end{equation}
We assume $\Delta_\mathrm{max}'>> \omega_\mathrm{r}=\Delta_\mathrm{min}'>\Gamma_2$. The first term corresponds to the typical  dispersive shift due to the interaction between an off-resonant TLS and a resonator. The second term is appreciable only under nonequilibrium conditions where \(\mathrm{d}\tilde{S}>0\) holds over a broad TLS spectral range. For the transformation from the summation to the integration in the second line and the calculation of the integration,
\begin{equation}
\begin{aligned}
&\underbrace{-
\int_{\Delta_{\mathrm{-max1}}}^{\Delta_\mathrm{-min1}}\mathrm{d}\Delta_{\mathrm{TLS}}
\frac{\Delta_{\mathrm{TLS}}}{\tilde{\Gamma}_2^2+\Delta_{\mathrm{TLS}}^2}}_{\Delta_\mathrm{TLS}<0}
\underbrace{-\int_{\Delta_{\mathrm{+min2}}}^{\Delta_\mathrm{+max2}}\mathrm{d}\Delta_{\mathrm{TLS}}
	\frac{\Delta_{\mathrm{TLS}}}{\tilde{\Gamma}_2^2+\Delta_{\mathrm{TLS}}^2}}_{\Delta_{\mathrm{TLS}}>0}
\\
&=+\int_{\Delta_{\mathrm{+min1}}}^{\Delta_\mathrm{+max1}}\mathrm{d}\Delta_{\mathrm{TLS}}
\frac{\Delta_{\mathrm{TLS}}}{\tilde{\Gamma}_2^2+\Delta_{\mathrm{TLS}}^2}
-\int_{\Delta_{\mathrm{+min2}}}^{\Delta_\mathrm{+max2}}\mathrm{d}\Delta_{\mathrm{TLS}}
\frac{\Delta_{\mathrm{TLS}}}{\tilde{\Gamma}_2^2+\Delta_{\mathrm{TLS}}^2}
\\
&=\mathrm{ln}\qty(\frac{\Delta_{\mathrm{max1}}}{\Delta_{\mathrm{min1}}})-\mathrm{ln}\qty(\frac{\Delta_{\mathrm{max2}}}{\Delta_{\mathrm{min2}}})
=\mathrm{ln}\qty(\frac{\Delta_\mathrm{max1}}{\Delta_{\mathrm{max2}}})
\rightarrow
\mathrm{ln}\qty(\frac{\Delta_\mathrm{max}'}{\Delta_{\mathrm{min}}'})
,
\end{aligned}
\end{equation}
is used assuming the lower cutoff detunings are the same $\Delta_{\mathrm{+min1}}=\Delta_{\mathrm{+min2}}>\tilde{\Gamma}_{2}$.

At 10--20 mK, most TLSs are in their ground state without optical illumination, producing a dispersive shift whose sign and magnitude depend on the detuning. Flipping high-energy ($\Delta_\mathrm{TLS}<0$) TLSs cancels the frequency-lowering dispersive contribution, resulting in a blue shift (frequency increase). Conversely, flipping low-energy ($\Delta_\mathrm{TLS}>0$) TLSs cancels the frequency-raising dispersive contribution, resulting in a red shift (frequency decrease). When $\omega_\mathrm{r}\sim10~\mathrm{GHz}$ and the upper cutoff for the relevant TLS frequencies satisfies $\Delta_{\mathrm{max}}'\gtrsim100~\mathrm{GHz}$, the number of high-energy TLSs within the integration window exceeds that of low-energy TLSs. Thus, on average, a net blue shift from the first term is expected under nonequilibrium excitation of TLSs.

\subsubsection{Effects of microwave-drive-induced saturation and TLSs' spectral diffusion}

We also consider the possibility that TLSs' spectral diffusion induced by TLS-TLS interactions can affect the resonator loss under the effect of TLS saturation. For TLSs with Gaussian spectral diffusion, $\sigma_\mathrm{sd}$, the total loss is given by

\begin{equation}\label{eq:spectral_diffusion_saturation}
\begin{aligned}
\delta\kappa_\mathrm{TLS,sd}
&
\simeq
-\hbar\rho_{\mathrm{TLS}}V_\mathrm{eff}
\int_{\Delta_\mathrm{min}}^{\Delta_\mathrm{max}}
\mathrm{d}\Delta_{\mathrm{TLS}}
\int_{\Delta_{\mathrm{TLS}}-\omega_\mathrm{cutoff}}^{\Delta_{\mathrm{TLS}}+\omega_\mathrm{cutoff}}\mathrm{d}\mu
\frac{2\tilde{g}_{\mathrm{\perp}}^2\tilde{\Gamma}_{2}}{\tilde{\Gamma}_{2}^2+\mu^2}
\frac{\tilde{S}}{1+
	\frac{n_\mathrm{cav}}{\tilde{n}_s}
	\frac{\tilde{\Gamma}_{2}^2}{\tilde{\Gamma}_{2}^2+\mu^2}
}
\frac{1}{\tilde{\sigma}_{\mathrm{sd}}\sqrt{2\pi}}\mathrm{exp}\qty(-\frac{(\mu-\Delta_{\mathrm{TLS}})^2}{2\tilde{\sigma}_{\mathrm{sd}}^2})
\\
&=-2\pi\hbar\rho_{\mathrm{TLS}}V_\mathrm{eff}\tilde{g}_\perp^2\tilde{S}/\sqrt{1+n_\mathrm{cav}/n_\mathrm{s}}\geqq0
,
\end{aligned}
\end{equation}
irrespective with the magnitude of $\sigma_\mathrm{sd}$, which coincides with the result obtained without spectral diffusion. Thus, Gaussian spectral diffusion by itself does not enhance the loss beyond the saturated-Lorentzian result.

\subsection*{Theoretical model for $\mathrm{d}\Delta(1/Q)/\mathrm{d}P_\mathrm{opt}$ and $\mathrm{d}(\Delta f_\mathrm{r}/f_\mathrm{r})/\mathrm{d}P_\mathrm{opt}$}
Based on Eqs. \eqref{eq:delta_kappa_tot}\eqref{eq:delta_f_tot}, we discuss the theoretical model describing the optically induced loss and frequency shift. First, since the experiments are performed in high-microwave-power regime, $\ev{\hat{\sigma}_z}\sim\tilde{S}\sim0$ for TLSs close to the resonance. Thus, we ignore the resonant loss term in $\delta\kappa_{\mathrm{TLS,total}}$ in the following analysis. In other words, we attribute the degradation in $Q$ to an increase in non-resonant loss. To explain the  optically induced effects, we assume the following processes are relevant:
\begin{enumerate}
	\item Optical illumination generates nonequilibrium high-energy phonons via recombination of quasiparticles in the superconductor and of electron–hole pairs in the silicon substrate.
	
	\item Nonequilibrium phonons flip TLS states in the relevant region via phonon absorption and emission.
	
	\item The region over which nonequilibrium phonons diffuse and affect TLS states broadens with optical power.
	
	\item Under interaction with nonequilibrium phonons, the TLS state can be described using single time- and spatially averaged parameters $\tilde{S}$ ($-1\le \tilde{S}\le 0$) and $\mathrm{d}\tilde{S}$.
\end{enumerate}

\subsubsection{Analytical model}
As mentioned in the main text, we make the following additional assumptions to model the laser-induced loss:
\begin{itemize}
	\item The nonequilibrium phonon diffusion length scales as $\xi P_\mathrm{opt}$, so that $A\xi P_\mathrm{opt}=V_\mathrm{eff}$.
\end{itemize}
Here, $A$ is the cross-sectional area of the region for consideration around the nanowire. Then, the optically induced change in $1/Q$ can be expressed as
\begin{equation}
\begin{aligned}
\Delta\qty(\frac{1}{Q})
=
\frac{\delta\kappa_\mathrm{TLS,total}}{\omega_\mathrm{r}}
&=
\frac{2\hbar\rho_{\mathrm{TLS}}A\xi
}{\omega_\mathrm{r}}
\frac{\tilde{\Gamma}_{1}\omega_\mathrm{r}}{\tilde{\Gamma}_{1}^2+\omega_\mathrm{r}^2}
\omega_\mathrm{max}\mathrm{d}\tilde{S}
\times
\tilde{g}_{\parallel}^2
\times P_\mathrm{opt}
\\
&\simeq2CK_{\parallel}\omega_\mathrm{r}\tilde{g}^2P_\mathrm{opt},
\end{aligned}
\end{equation}
where $C=\hbar\rho_{\mathrm{TLS}}A\xi/\omega_\mathrm{r}$, and $K_{\parallel}=\tilde{\Gamma}_1\omega_\mathrm{max}\mathrm{d}\tilde{S}/(\tilde{\Gamma}_1^2+\omega_\mathrm{r}^2)$.
Meanwhile, $\Delta f_\mathrm{r}/f_\mathrm{r}$ can be written as
\begin{equation}\label{eq:analytical_f}
	\begin{aligned}
		\frac{\Delta f_\mathrm{r}}{f_\mathrm{r}}
		=
		\frac{\delta \omega_\mathrm{TLS,total}}{\omega_\mathrm{r}}
		&=
		\qty[
		\frac{\hbar\rho_{\mathrm{TLS}}A\xi
		}{\omega_\mathrm{r}}
		\mathrm{ln}\qty(\frac{\Delta_\mathrm{max}'}{\Delta_{\mathrm{min}}'})
		(1+\tilde{S})
		\times\tilde{g}_{\perp}^2
		-\frac{\hbar\rho_{\mathrm{TLS}}A\xi
		}{\omega_\mathrm{r}}
		\frac{\tilde{\Gamma}_{1}^2}{\tilde{\Gamma}_{1}^2+\omega_\mathrm{r}^2}
		\omega_\mathrm{max}\mathrm{d}\tilde{S}	
		\times\tilde{g}_{\parallel}^2
		]
		\times P_\mathrm{opt}
		\\
		&\simeq
		C(K_\perp-\tilde{\Gamma}_1K_\parallel)\tilde{g}^2P_\mathrm{opt}
		,
	\end{aligned}
\end{equation}
where $K_\perp=(1+\tilde{S})\mathrm{ln}(\Delta_\mathrm{max}'/\Delta_{\mathrm{min}}')$.
Here, $(1+\tilde{S})$ in the first term differs from Eq.~\eqref{eq:delta_f_tot} because we assume that optical illumination cancels the baseline dispersive shift by increasing the excited-state population.

For the calculation in the main text, $\omega_\mathrm{r}/2\pi=7$ GHz, $\tilde{S}=0$, $2\pi\mathrm{d}\tilde{S}=$ 1/400 MHz$^{-1}$,  $\tilde{\Gamma}_1/2\pi=$ 16 MHz ($T_1\simeq10$ ns), 
$\rho_\mathrm{TLS}=1\times10^{45}$ J$^{-1}$m$^{-3}$, $A=tw$ (TLS layer thickness $t=2$ nm; effective width $w=500$ nm), 
$\Delta_\mathrm{max}/2\pi=\omega_\mathrm{max}/2\pi=1000$ GHz, and
$\Delta_\mathrm{min}/2\pi=\omega_\mathrm{r}/2\pi=7$ GHz are used. Also we assume $\tilde{g}_\perp=\tilde{g}_\parallel$. 
Table \ref{tab:sammary_para_TLS_calc} summarizes the parameters. 

\begin{figure}[tb]
	\centering
	\includegraphics[width=18cm,clip]{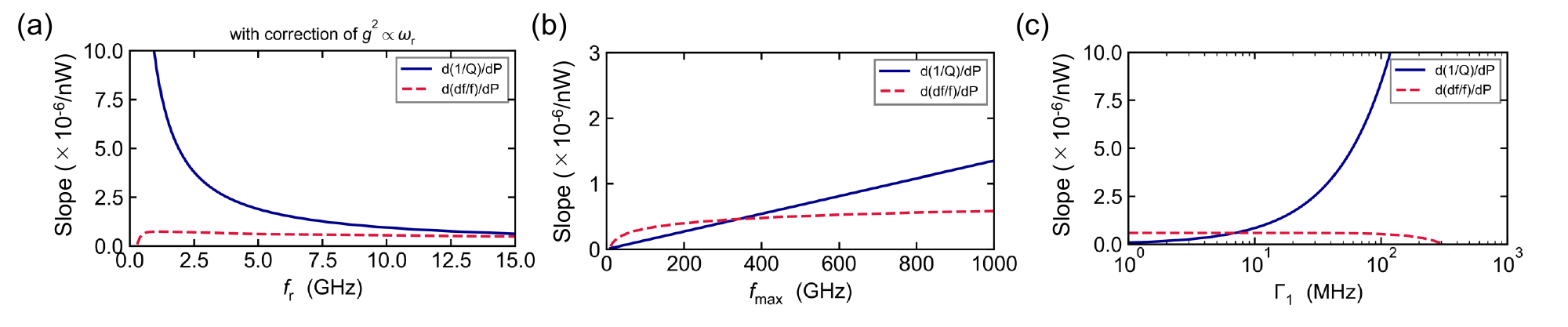}
	\caption{$\mathrm{d}\Delta(1/Q)/\mathrm{d}P_\mathrm{opt}$ and $\mathrm{d}(\Delta f_\mathrm{r}/f_\mathrm{r})/\mathrm{d}P_\mathrm{opt}$ as a function of (a) the resonance frequency, $f_\mathrm{r}$, (b) the cutoff frequency for integration, $f_\mathrm{max}$, (c) the relaxation time, $\tilde{\Gamma}_{1}$. For the calculation of $f_\mathrm{r}$ dependence, the result is normalized by $\omega_\mathrm{r}$ to incorporate the frequency dependence of $g^2\propto\omega_\mathrm{r}$. }
	\label{fig:Qinv_df_para_dependence}
\end{figure}

We also examine parameter dependence of Eqs.  \eqref{eq:delta_kappa_tot}\eqref{eq:delta_f_tot} as shown in  Fig.\ref{fig:Qinv_df_para_dependence}. 
Fig.\ref{fig:Qinv_df_para_dependence} (a) shows the resonance frequency dependence. Since $g^2\propto f_\mathrm{r}$, we account for this by setting $\tilde{g}(\omega_\mathrm{r})=\tilde{g}_0\sqrt{f{_r}/7~\mathrm{GHz}}$ ($\tilde{g}_0/2\pi=5~\mathrm{MHz}$). Then, $\mathrm{d}(1/Q)/\mathrm{d}P_\mathrm{opt}$ depends on $\omega_\mathrm{r}$ as $\propto1/\omega_\mathrm{r}$, whereas $\mathrm{d}(\Delta f_\mathrm{r}/f_\mathrm{r})/\mathrm{d}P_\mathrm{opt}$ shows only weak dependence on $\omega_\mathrm{r}$ except below $1~\mathrm{GHz}$. For dependence on the cutoff frequency, $\omega_\mathrm{max}=\Delta_{\mathrm{max}}$, Fig. \ref{fig:Qinv_df_para_dependence} (b) shows that $\mathrm{d}(1/Q)/\mathrm{d}P_\mathrm{opt}\propto \omega_\mathrm{max}$ and $\mathrm{d}(\Delta f_\mathrm{r}/f_\mathrm{r})/\mathrm{d}P_\mathrm{opt}\propto\mathrm{ln}(\omega_\mathrm{max})$. As a function of $\Gamma_{1}$, $\mathrm{d}(1/Q)/\mathrm{d}P_\mathrm{opt}$ increases approximately linearly with $\Gamma_1$, whereas $\mathrm{d}(\Delta f_\mathrm{r}/f_\mathrm{r})/\mathrm{d}P_\mathrm{opt}$ decreases roughly quadratically with $\Gamma_1$.

\begin{center}
\begin{threeparttable}[tb]
	\caption{Summary of parameters used for the theoretical calculations.}
	\label{tab:sammary_para_TLS_calc}
	\begin{tabular}{l l c l l}
		\hline
		Nonequilibrium parameter ($-1\leqq\ev{\hat{\sigma}_z}_0\leqq0$)
		& $\tilde{S}$
		& 0
		& 
		&	
		\\
		Nonequilibrium parameter ($0\leqq\mathrm{d}(\ev{\hat{\sigma}_z}_0)
		/\mathrm{d}\omega|_{\omega=\omega_\mathrm{TLS}}$)
		&$\mathrm{d}\tilde{S}\times2\pi$
		& 1/400		
		& MHz$^{-1}\tnote{*}$
		&
		\\
		Resonator frequency & $\omega_\mathrm{r}/2\pi$ & 7 & GHz &
		\\
		TLS relaxation time & $\tilde{\Gamma}_1/2\pi$ &16&MHz\tnote{**}&
		\cite{Shalibo2010,Lisenfeld2019}
		\\
		TLS density of states & $\rho_{\mathrm{TLS}}$ & $1\times10^{45}$ & J$^{-1}$\,m$^{-3}$ &
		\cite{Constantin2007,Skacel2014,Burnett2016,Lisenfeld2019}
		  \\
		TLS layer thickness ($\simeq$ SiO$_2$ thickness) & $t$ & 2 & nm &  \\
		Relevant width around the nanowire & $w$ & 500 & nm &  \\
		Cross-sectional area of the region considered & $A=t\,w$ & $1000$ & nm$^2$ &  \\
		Cutoff detuning for the calculation of the loss&
		$\omega_\mathrm{max}/2\pi$&1000& GHz&
		\\
		Cutoff detuning for the calculation of the frequency shift & $\Delta_{\max}/2\pi$ & 1000 & GHz &  \\
		Cutoff detuning for the calculation of the frequency shift & $\Delta_{\min}/2\pi$ & 7 & GHz &  \\
		TLS-resonator coupling in the relevant area & $\tilde{g}_\perp/2\pi=\tilde{g}_\parallel/2\pi$=$\tilde{g}/2\pi$ & 0--10 & MHz &\cite{Kristen2023}
		\\
		Scaling parameter for phonon diffusion length & $\xi$ & 5--250 & m/W &
		\\
		\hline
	\end{tabular}
\begin{tablenotes}
	\item[*] $\mathrm{d}\tilde{S}\times2\pi\sim1/400$ MHz$^{-1}$ corresponds to $T\simeq$ 20 mK if we use the equation based on equilibrium temperature,  $[1-\mathrm{tanh}^2(\hbar\omega_\mathrm{TLS}/k_BT)]/k_BT=\mathrm{d}\tilde{S}$.
	\item[**]
	$\Gamma_1\sim16$ MHz corresponds to $T_1$ of $\sim$ 10 ns. Although $T_1$ of typical TLSs is 100--10000 ns at $<$ 10 GHz, assuming that high-frequency TLSs have shorter $T_1$ due to an increase in the phonon DOS, shorter $T_1$ is plausible.
\end{tablenotes}
\end{threeparttable}
\end{center}


\subsubsection{Monte Carlo model}

For the numerical simulations, we add an additional assumption: 
\begin{itemize}
	\item The scaling kernel function, $K(x,P_\mathrm{opt})$, weights the contribution from each TLS,
\end{itemize}
to smooth the $\Delta(1/Q) (P_\mathrm{opt})$ and $\Delta f_\mathrm{r}/f_\mathrm{r}(P_\mathrm{opt})$ curves. Based on this assumption, $\Delta(1/Q)(P_\mathrm{opt})$ and $\Delta f_\mathrm{r}/f_\mathrm{r}(P_\mathrm{opt})$ can be written as the sum of weighted contributions from each TLS as
\begin{equation}\label{eq:monte-carlo-Q}
\Delta\qty(\frac{1}{Q})
=\frac{1}{\omega_\mathrm{r}}
\sum_{i} K(x_i,P_\mathrm{opt})\times
\mathrm{d}S_i
\frac{2\Gamma_{i,1}\omega_\mathrm{r}}{\Gamma_{i,1}^2+\omega_\mathrm{r}^2}
 g_{i,\parallel}^2
,
\end{equation}
and
\begin{equation}\label{eq:monte-carlo-f}
	\frac{\Delta f_\mathrm{r}}{f_\mathrm{r}}
	=\frac{1}{\omega_\mathrm{r}}
	\sum_{i} K(x_i,P_\mathrm{opt})\times	
	\qty[
	(1+S_i)
	\frac{\Delta_{i,\mathrm{TLS}}}{\Gamma_{i,2}^2+\Delta_{i,\mathrm{TLS}}^2}g_{i,\perp}^2
	-
	\mathrm{d}S_i
	\frac{\Gamma_{i,1}^2}{\Gamma_{i,1}^2+\omega_\mathrm{r}^2}
	g_{i,\parallel}^2
	],
\end{equation}
where 
\begin{equation}
	K(x_i,P_\mathrm{opt})=\frac{1}{2}\qty[\mathrm{tanh}\qty(\frac{x_i+\xi P_\mathrm{opt}/2}{l_\mathrm{edge}})-\mathrm{tanh}\qty(\frac{x_i-\xi P_\mathrm{opt}/2}{l_\mathrm{edge}})],
\end{equation}
where $x_i$ is the distance from the laser spot, $l_\mathrm{edge}$ sets the cutoff length scale of the hyperbolic tangent, and $i$ indexes a TLS within the considered frequency–position window, randomly generated in the simulation.
$l_\mathrm{edge} = 10$ $\mu$m is used. The maximum cutoff frequencies for generating TLSs are the same as those used for the analytical calculation, 1000 GHz. The minimum cutoff frequency for the relevant TLS generation is set to  $|\Delta_\mathrm{TLS}/2\pi|\geqq100$ MHz to avoid the large fluctuation coming from nearly-resonant TLSs. We set $g_{i,\perp}=g_{i,\parallel}$ and $\Gamma_{i,1}=\Gamma_{i,2}$ for simplicity, assigning them probabilistically from Gaussian distributions centered at $\tilde{g}$ and $\tilde{\Gamma}_1$, respectively. The standard deviations of $\tilde{g}$ and $\tilde{\Gamma}_1$ are set to their central values divided by $2\sqrt{2\ln 2}$. Each $S_i$ is drawn from a Gaussian with mean $0$ and standard deviation $0.35$.

\begin{figure}[tb]
	\centering
	\includegraphics[width=16cm,clip]{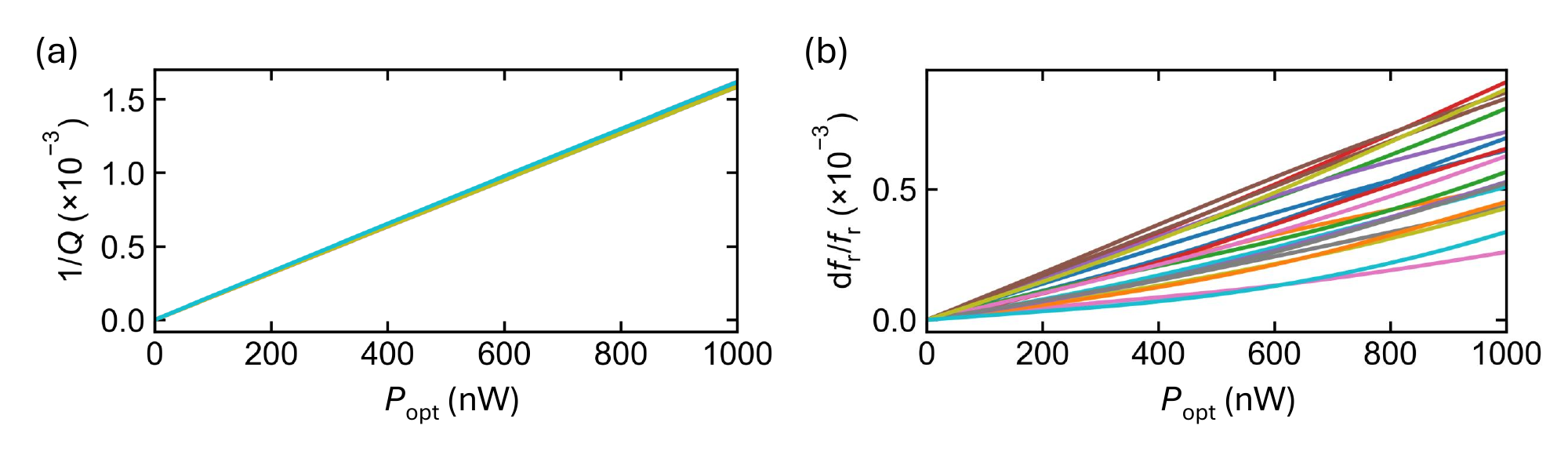}
	\caption{Monte Carlo simulation of $\Delta(1/Q)$ and $\Delta f_\mathrm{r}/f_\mathrm{r}$ for parameters in Table \ref{tab:sammary_para}. $\tilde{g}/2\pi$ = 5 MHz and $\xi$ = 50 m/W. 
	}
	\label{fig:monte_carlo}
\end{figure}

Fig. \ref{fig:monte_carlo} show Monte Carlo simulations of $1/Q$ and $\Delta f_\mathrm{r}/f_\mathrm{r}$ as a function of $P_\mathrm{opt}$. The mean over 100 trials yields $\mathrm{d}\Delta(1/Q)/\mathrm{d}P_\mathrm{opt}$ and $\mathrm{d}(\Delta f_\mathrm{r}/f_\mathrm{r})/\mathrm{d}P_\mathrm{opt}$ values in good agreement with the analytical model. Whereas $1/Q$ does not show differences in each trial, $\Delta f_\mathrm{r}/f_\mathrm{r}$ shows a large standard deviation across trials. This is because $\Delta f_\mathrm{r}/f_\mathrm{r}$ is sensitive to the number of TLSs with small detuning. On the other hand, $1/Q$ does not depend on the detuning, which is characteristic of the  longitudinal coupling.

The dependence of $\mathrm{d}(\Delta f_\mathrm{r}/f_\mathrm{r}) /\mathrm{d}P_\mathrm{opt}$ on the random TLS frequency and positional distribution can partially explain the deviation from the global trend in the third and fifth modes at Area 4. Based on the simulated local current density, the red (blue) shift is expected for the third (fifth) mode. However, in practice, the subtle (red) shift is observed for the third (fifth) mode. If the local density of relevant TLS is higher at frequencies between the third and fifth modes than elsewhere, the blue-shift contribution (for the third mode) and the red-shift contribution (for the fifth mode) become larger, accounting for the observed deviations.

\subsection*{The nonlinear behavior in $\Delta f_\mathrm{r}/f_\mathrm{r}$}

\begin{figure}
	\centering
	\includegraphics[width=14cm,clip]{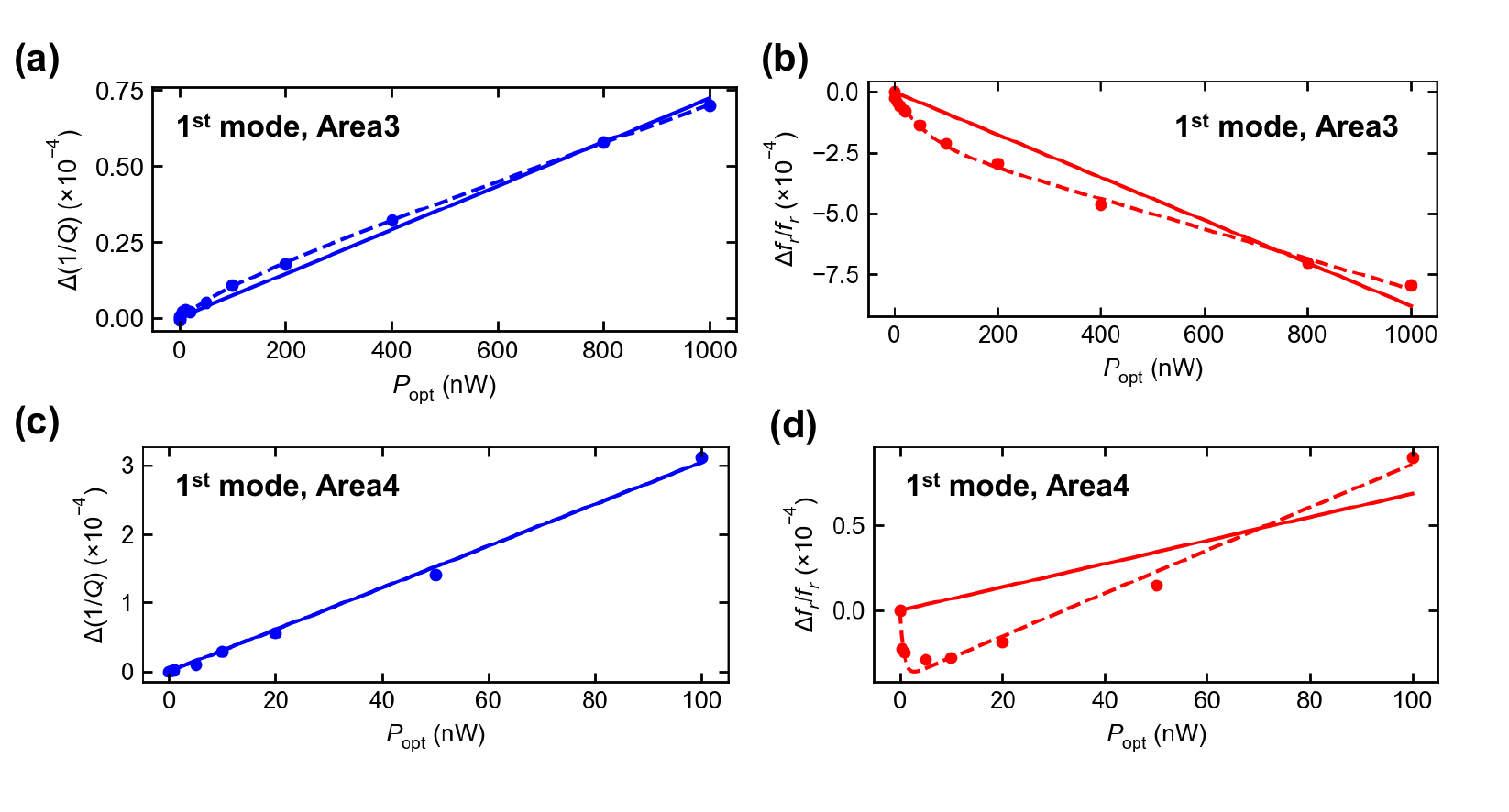}
	\caption{ (a) $\Delta (1/Q)$ and (b) $\Delta f_\mathrm{r}/f_\mathrm{r}$ of the first mode at Area 3. (a) $\Delta (1/Q)$ and (b) $\Delta f_\mathrm{r}/f_\mathrm{r}$ of the first mode at Area 4. The red (blue) lines are the linear fit, while the dotted lines are the linear fit with the saturating term. }
	\label{fig:linear_vs_saturation}
\end{figure}

We discuss the nonlinear behavior in $\Delta f_\mathrm{r}/f_\mathrm{r}$, which is not treated in the main text. Figure~\ref{fig:linear_vs_saturation} shows $\Delta(1/Q)$ and $\Delta f_\mathrm{r}/f_\mathrm{r}$ for the first mode in Area~3 and Area~4. The data are fitted with a linear model and a linear-plus-saturation model:
$
\Delta(1/Q)=\gamma_1 P_\mathrm{opt}+\gamma_2\!\left[1-\exp\!\left(-\gamma_3 P_\mathrm{opt}\right)\right],\qquad
\Delta f_\mathrm{r}/f_\mathrm{r}=\delta_1 P_\mathrm{opt}+\delta_2\!\left[1-\exp\!\left(-\delta_3 P_\mathrm{opt}\right)\right].
$
The saturating behavior is clearly visible in $\Delta f_\mathrm{r}/f_\mathrm{r}$ in both areas. In Area~3, saturation appears at high optical powers ($\gtrsim 100~\mathrm{nW}$), whereas in Area~4 it emerges at low powers ($\lesssim 10~\mathrm{nW}$). Additionally, for $\Delta(1/Q)$ in Area~3, the saturating model reproduces the data slightly better than a simple linear fit.

A possible explanation for the moderate saturation at high optical powers is saturation of the quasiparticle recombination time. Within a recombination-limited picture, the optically induced quasiparticle density obeys
$
\mathrm{d}n_\mathrm{qp}/\mathrm{d}P_\mathrm{opt}\propto \bigl(1+\eta P_\mathrm{opt}\bigr)^{-1/2},
$
where $\eta$ is a constant~\cite{Day2024}. In the low-power limit, $\Delta n_\mathrm{qp}\propto P_\mathrm{opt}$, consistent with the main-text assumption. Since $\Delta(1/Q)\propto\Delta n_\mathrm{qp}$ and $\Delta f_\mathrm{r}/f_\mathrm{r}\propto\Delta n_\mathrm{qp}$, both quantities can exhibit sublinear, saturating shifts at high optical powers.

For the sharp low-power saturation in $\Delta f_\mathrm{r}/f_\mathrm{r}$, no concomitant saturation is observed in $\Delta(1/Q)$, in contrast to the moderate high-power case. Because $n_\mathrm{qp}$ affects both $\Delta(1/Q)$ and $\Delta f_\mathrm{r}/f_\mathrm{r}$, a quasiparticle-based origin can be excluded. Additionally, the saturating shift in $\Delta f_\mathrm{r}/f_\mathrm{r}$ is consistently downward across other modes and areas. Furthermore, phonon-mediated TLS flipping cannot explain this low-power saturation. As discussed above, the phonon-activated region grows with optical power (diffusion over several to tens of micrometers), which would cause a non-saturating shift. Based on these observations, an optically active local TLS with a transition frequency below $\omega_\mathrm{r}$ can explain the low-power saturation in $\Delta f_\mathrm{r}/f_\mathrm{r}$. If laser illumination promotes such a TLS to an optically excited state or a different charge configuration, its population imbalance changes and its dispersive contribution is effectively turned off. Moreover, if the optically active TLS has a microwave transition with $\omega_\mathrm{TLS}<\omega_\mathrm{r}$, removing its frequency-raising (dispersive) contribution yields a downward shift. Because the laser spot is $\sim\!1~\mu\mathrm{m}^2$, the optically activated region is similarly limited, in contrast to the phonon-activated region, which scales with optical power. Thus, this frequency-lowering shift saturates once all relevant TLSs within the illuminated area are activated. Although the existence of such TLSs is not guaranteed, hydroxyl-related defects are known to exhibit both optical and microwave transitions~\cite{Plotnichenko2000,Khalil2013}.

\subsection*{Consideration of the contribution from an increase in equilibrium temperature}

In the literature \cite{Wang2013a,Janssen2014}, an equivalence between optical power and temperature has been discussed.
However, in this paper we employ a TLS-based model that does not invoke changes in the equilibrium temperature.
In this section, we assess the possible contribution of temperature-dependent TLS-permittivity shifts.

We estimate the expected change in $\Delta f_\mathrm{r}/f_\mathrm{r}$ caused by increases in the equilibrium temperature and compare it with experiment.
From Eq.~\ref{eq:freq_shift_final}, which is based on temperature-dependent TLS permittivity, an increase in temperature produces a saturating downward shift at low temperatures ($T<hf_\mathrm{r}/k_B$) and a saturating upward shift near and above $T\sim hf_\mathrm{r}/k_B$ (up to $\lesssim 1~\mathrm{K}$).
At higher temperatures ($T>1~\mathrm{K}$), the quasiparticle-induced downward shift dominates.
Measurements of $\Delta f_\mathrm{r}(T)/f_\mathrm{r}$ show TLS-induced shifts ranging from approximately $-2\times10^{-5}$ (downward) to $+4\times10^{-5}$ (upward) (Fig.~\ref{fig:temp-freq}).
Assuming that only a few tens of $\mu\mathrm{m}^2$ within the $1.5$-mm-long nanowire are heated by optical illumination, the participation ratio is at most $0.01$–$0.1$.
The expected frequency shift is therefore at most $\sim 4\times10^{-6}$ (i.e., $4\times10^{-5}\times 0.1$).
By contrast, laser irradiation produces a frequency shift exceeding $1\times10^{-4}$.
Thus, the observed frequency shift cannot be explained solely by an increase in the equilibrium temperature.

For $\Delta(1/Q)$, according to the standard tunneling model, increasing temperature reduces TLS-induced loss via saturation (Eq.~\ref{eq:TLS-loss-temp}).
However, we observe an increase in TLS-induced loss with increasing optical power.
Additionally, from the microwave-power-dependence measurement (Fig.~\ref{fig:mw_power_Q}), the TLS loss without laser illumination is $\sim 2\times10^{-5}$, which is already saturated at the high microwave drive powers used in the experiments.
Thus, again, the laser-induced increases in $\Delta(1/Q)$ ($>1\times10^{-4}$) cannot be explained solely within an equilibrium-temperature model.


\clearpage